\documentclass{aa}    
\usepackage{txfonts}
\usepackage{longtable}  
\usepackage{natbib}
\usepackage{graphicx}

\bibpunct[, ]{(}{)}{,}{a}{}{,}

\usepackage{color}

\newcommand{\Eexc}{$E_{\rm exc}$}
\newcommand{\Teff}{T_{\rm eff}}
\newcommand{\logg}{\rm log~ g}
\newcommand{\kms}{km\,s$^{-1}$}
\newcommand{\eps}[1]{\log\varepsilon_{\rm #1}}
\newcommand{\kH}{$S_{\!\rm H}$}    %%% Note negative spaces!
\newcommand{\eu}[5]{\mbox{$#1\,^#2{\rm #3}^{#4}_{\rm #5}$}}

\begin{document}

\title{Influence of inelastic collisions with hydrogen atoms on the non-LTE line formation for \ion{Fe}{i} and \ion{Fe}{ii} in the 1D model atmospheres of late-type stars}

\author{
L. Mashonkina\inst{1} \and 
T. Sitnova\inst{1} \and
S.~A. Yakovleva\inst{2} \and
A.~K. Belyaev\inst{2}
}

\offprints{L. Mashonkina; \email{lima@inasan.ru}}
\institute{
Institute of Astronomy, Russian Academy of Sciences, RU-119017 Moscow, 
     Russia \\ \email{lima@inasan.ru}
\and Department of Theoretical Physics and Astronomy, Herzen University, St. Petersburg 191186, Russia
}

\date{Received  / Accepted }

\abstract{
Iron plays a crucial role in studies of late-type stars. In their atmospheres, neutral iron is the minority species and lines of \ion{Fe}{i} are subject to the departures from LTE. In contrast, one believes that LTE is a realistic approximation for \ion{Fe}{ii} lines. The main source of the uncertainties in the non-local thermodynamic equilibrium (non-LTE = NLTE) calculations for cool atmospheres is a treatment of inelastic collisions with hydrogen atoms.}
{We investigate the effect of \ion{Fe}{i} + \ion{H}{i} and \ion{Fe}{ii} + \ion{H}{i} collisions and their different treatment on the \ion{Fe}{i}/\ion{Fe}{ii} ionisation equilibrium and iron abundance determinations for Galactic halo benchmark stars HD~84937, HD~122563, and HD~140283 and a sample of 38 very metal-poor (VMP) giants in the dwarf galaxies with well known distances.}
{We perform the NLTE calculations for \ion{Fe}{i} - \ion{Fe}{ii} with applying quantum-mechanical rate coefficients for collisions with \ion{H}{i} from Barklem (2018, B18), Yakovleva, Belyaev, and Kraemer (2018, YBK18), and Yakovleva, Belyaev, and Kraemer (2019, YBK19).} 
{We find that collisions with \ion{H}{i} serve as efficient thermalisation processes for \ion{Fe}{ii}, such that the NLTE abundance corrections for \ion{Fe}{ii} lines do not exceed 0.02~dex, in absolute value, at [Fe/H] $\gtrsim -3$ and reach +0.06~dex at [Fe/H] $\sim -4$. For a given star, different treatments of \ion{Fe}{i} + \ion{H}{i} collisions by 
 B18 and YBK18 lead to similar average NLTE abundances from the \ion{Fe}{i} lines, although there exist discrepancies in the NLTE abundance corrections for individual lines. 
% We note, in particular, the lines arising from the \ion{Fe}{i} \eu{z}{5}{D}{\circ}{} level. 
With using quantum-mechanical collisional data and the Gaia based surface gravity, we obtain consistent abundances from the two ionisation stages, \ion{Fe}{i} and \ion{Fe}{ii}, for a red giant HD~122563.  
For a turn-off star HD~84937 and a subgiant HD~140283, we analyse the iron lines in the visible and the ultra-violet (UV, 1968 to 2990~\AA) range. For either \ion{Fe}{i} or \ion{Fe}{ii}, abundances from the visible and UV lines are found to be consistent in each star. The NLTE abundances from the two ionisation stages agree within 0.10~dex, when using the YBK18 data, and 0.13~dex in case of B18. 
%When using all the lines, an abundance discrepancy betweendoes not exceed 0.10~dex in the NLTE calculations with  
The \ion{Fe}{i}/\ion{Fe}{ii} ionisation equilibrium is achieved for each star of our stellar sample in the dwarf galaxies, with the exception of stars at [Fe/H] $\lesssim -3.7$.
%We discuss the effect of the uncertainties in collisional data on the derived iron abundances.
}
{}
% in not only visible, but also ultra-violet (UV, 1968 to 2990~\AA)
 

\titlerunning{NLTE line formation for \ion{Fe}{i} and \ion{Fe}{ii} in atmospheres of late-type stars }
\authorrunning{Mashonkina et al.}

\maketitle

\section{Introduction} \label{sec:intro}

Iron is of extreme importance for studies of cool stars. It serves as a reference element for all astronomical research related to stellar nucleosynthesis and chemical evolution of the Galaxy thanks to the many lines in the visible spectrum, which are easy to detect even in ultra ([Fe/H]\footnote{In the classical notation, where [X/H] = $\log(N_{\rm X}/N_{\rm H})_{star} - \log(N_{\rm X}/N_{\rm H})_{Sun}$.} $< -4$, UMP) and hyper ([Fe/H] $< -5$) metal-poor stars. Lines of iron are used to determine basic stellar atmosphere parameters, that is, the effective temperature, $\Teff$, from the excitation equilibrium of \ion{Fe}{i} and the surface gravity, $\logg$, from the ionisation equilibrium between \ion{Fe}{i} and \ion{Fe}{ii}. Neutral iron is a minority species in stellar atmospheres with $\Teff > 4500$~K. Therefore, its statistical equilibrium (SE) can easily deviate from thermodynamic equilibrium due to deviations of the mean intensity of ionising radiation from the Planck function, and the theoretical spectra need to be modelled based on the non-local thermodynamic equilibrium (non-LTE = NLTE) line formation. 

For last half century, the model atoms for \ion{Fe}{i} (and \ion{Fe}{ii}) were developed in several
%the problem of  for iron in the atmospheres of the Sun and cool stars was considered in many
studies \citep[see, for example][and references therein]{1971PASJ...23..217T,1985Ap.....22..203B,1986A&A...165..170G,2001A&A...366..981G,2001ApJ...550..970S}. However, the results obtained for the populations of high-excitation levels of \ion{Fe}{i} were not always convincing \citep[for a discussion, see, for example][]{2003A&A...407..691K}. 
A step forward in improving the SE calculations for \ion{Fe}{i} - \ion{Fe}{ii} was made by \citet{mash_fe}, who included in the model atom the \ion{Fe}{i} energy levels from not only the laboratory measurements, but also atomic structure calculations, in total, about 3~000 levels. The predicted high-excitation levels of \ion{Fe}{i} nearly do not contribute to the total number density of iron, however, they play an important role in providing close collisional coupling of \ion{Fe}{i} to the large continuum electron reservoir that reduces the NLTE effects for lines of \ion{Fe}{i}. Similar approach was implemented later by \citet{Bergemann_fe_nlte}. The model atom of \citet{mash_fe} was applied to determine atmospheric parameters and iron abundances of extended stellar samples \citep[for example, ][]{2015ApJ...808..148S,lick_paperII,dsph_parameters,2017A&A...608A..89M}. The model atom of \citet{Bergemann_fe_nlte} was used to compute the grids of the NLTE abundance corrections \citep{2012MNRAS.427...50L}, which have wide applications \citep[for example,][]{Ruchti2013,Bensby2014,Bergemann2014A&A...565A..89B}.

The need for a new analysis of \ion{Fe}{i} - \ion{Fe}{ii} is motivated by the recent quantum-mechanical calculations of \citet{2018A&A...612A..90B} and \citet{2018CP....515..369Y} for inelastic \ion{Fe}{i} + \ion{H}{i} collisions and \citet{2019MNRAS.483.5105Y} for \ion{Fe}{ii} + \ion{H}{i} collisions. Until recent time, the treatment of poorly known inelastic collisions with hydrogen atoms was the main source of the uncertainties in the NLTE results. Previous NLTE studies of \ion{Fe}{i}-\ion{Fe}{ii} relied on \ion{H}{i} collision rates from a rough theoretical approximation of \citet{Drawin1968,Drawin1969}, as implemented by \citet{Steenbock1984}, and these rates were scaled by a factor \kH, which was constrained empirically. For example, \citet{mash_fe} estimated \kH\ = 0.1 from inspection of different influences of \ion{H}{i} collisions on the \ion{Fe}{i} and \ion{Fe}{ii} lines in the five benchmark stars with well-determined stellar parameters. Using nearly the same stellar sample, \citet{Bergemann_fe_nlte} recommended \kH\ = 1, which was adopted in  computations of the NLTE grids by \citet{2012MNRAS.427...50L}. \citet{2015ApJ...808..148S} and \citet{dsph_parameters} estimated \kH\ = 0.5 from independent analyses of the two extended stellar samples, namely, nearby dwarfs with accurate {\sc Hipparcos} parallaxes available and very metal-poor (VMP, [Fe/H] $< -2$) giants in the dwarf galaxies with known distances. 
 Quantum-mechanical rate coefficients for the \ion{Fe}{i} + \ion{H}{i} collisions were used  
 in only very few papers: \citet{2017MNRAS.468.4311L} and \citet{2016MNRAS.463.1518A} applied the data from a previous set of calculations of \citet{2016PhRvA..93d2705B}.

This study aims to investigate an influence of using accurate data on \ion{Fe}{i} + \ion{H}{i} and \ion{Fe}{ii} + \ion{H}{i} collisions on the \ion{Fe}{i} /\ion{Fe}{ii} ionisation equilibrium and 
iron abundance determinations for metal-poor stars. The model atom from \citet{mash_fe} is taken as a basic model and is updated for calculations of collisional rates. We use the three Galactic halo benchmark stars with well determined atmospheric parameters and a sample of the 38 VMP giants in the dwarf galaxies with known distances. We want to inspect the NLTE effects for lines of \ion{Fe}{ii}. Based on the available NLTE calculations \citep[for example,][]{Gratton1999,2001A&A...366..981G,mash_fe,Bergemann_fe_nlte}, one commonly believes that LTE is a realistic approximation for \ion{Fe}{ii} lines. However, all the cited NLTE studies used the Drawinian rates to take into account collisions with \ion{H}{i} in the SE calculations. How do the NLTE results for \ion{Fe}{ii} change, when using accurate data on \ion{Fe}{ii} + \ion{H}{i} collisions?

The paper is organised as follows. Section\,\ref{Sect:NLTE} describes the updated model atom of \ion{Fe}{i} -\ion{Fe}{ii} and the effects of using accurate collisional data on the SE of iron in the VMP atmospheres. In Section\,\ref{sec:stars} we determine abundances from lines of \ion{Fe}{i} and \ion{Fe}{ii} and inspect the \ion{Fe}{i} /\ion{Fe}{ii} ionisation equilibrium in the sample of  VMP stars. 
%The remaining uncertainties in the NLTE calculations for \ion{Fe}{i} -\ion{Fe}{ii} are discussed in Sect.\,\ref{Sect:uncertainties}. 
Section\,\ref{Sect:comparison} compares the obtained results with the literature data. Our recommendations and conclusions are given in Sect.\,\ref{Sect:Conclusions}.

% and they were widely applied in stellar parameter and abundance analyses \citep[see][for references]{aspl05}.

%\ion{Fe}{i} \citet[][hereafter, YBK18]{2018CP....515..369Y}
%\ion{Fe}{ii} \citet[][hereafter, YBK19]{2019MNRAS.483.5105Y}

\section{Method of NLTE calculations for \ion{Fe}{i}-\ion{Fe}{ii}} \label{Sect:NLTE}

%In this section, we describe updates of the model atom  \ion{Fe}{i} - \ion{Fe}{ii} and inspect the influence of inelastic collisions with \ion{H}{i} on the SE of iron. 
The coupled radiative transfer and SE equations are solved with the {\sc DETAIL} code
\citep{detail}.
%based on the accelerated lambda iteration method \citep[recipe of][]{rh91,rh92}. 
The opacity package in {\sc DETAIL} was updated as described by \citet{mash_fe}. This research uses the MARCS homogeneous plane-parallel model atmospheres with standard abundances \citep{Gustafssonetal:2008} available on the MARCS website\footnote{\tt
  http://marcs.astro.uu.se}. They were interpolated at the necessary $\Teff$,
$\logg$, and iron abundance [Fe/H], using the FORTRAN-based routine written by Thomas
Masseron that is available on the same website. For calculations with {\sc DETAIL}, the models were converted to the {\sc MAFAGS} format by applying the routines and input data from the {\sc MAFAGS-OS} code \citep{Grupp2009}.

 \begin{figure*}
% \begin{minipage}{190mm}
 \begin{center}
  \resizebox{90mm}{!}{\includegraphics{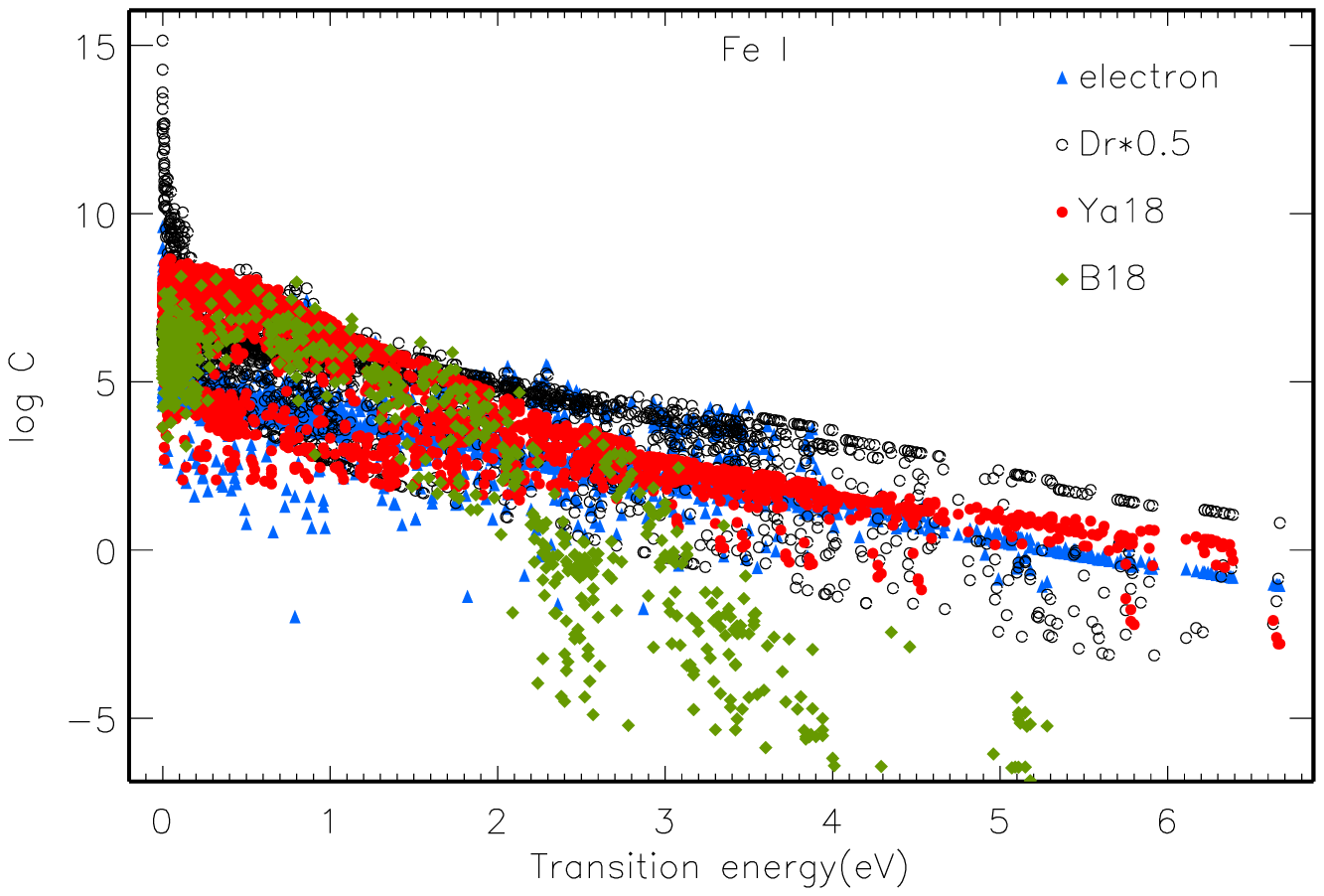}}
  \resizebox{90mm}{!}{\includegraphics{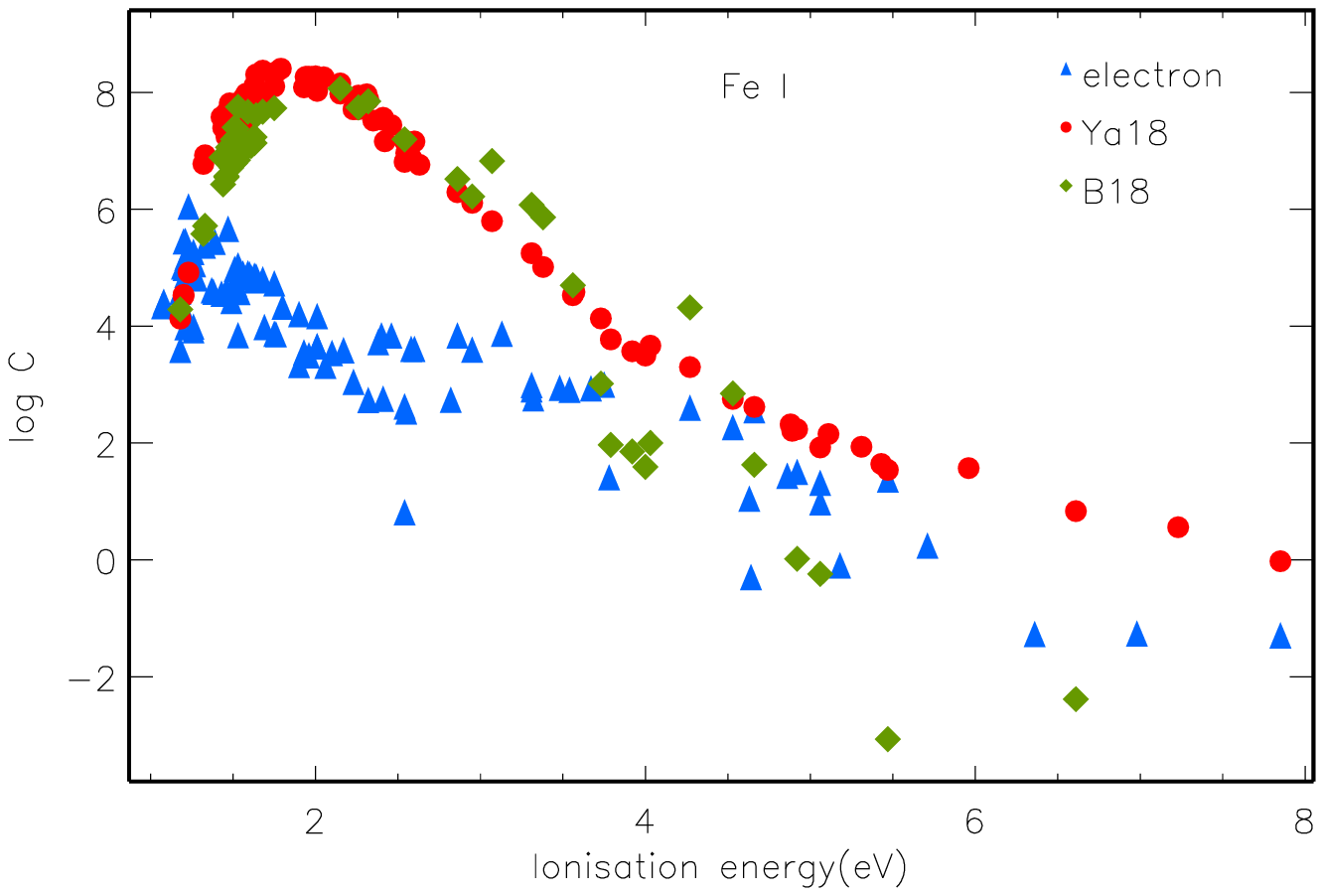}}
  \caption{Left panel: \ion{Fe}{i} excitation rates (in s$^{-1}$), log~C, for electron impact (triangles) compared with the rates for \ion{H}{i} collisions from calculations of YBK18 (filled circles) and B18 (rhombi) and compared with the scaled (\kH\ = 0.5) Drawinian rates (open circles). Right panel: rates, log~C, of the processes \ion{Fe}{i} + ${\rm e^-} \rightarrow$ \ion{Fe}{ii} + $2{\rm e^-}$ and \ion{Fe}{i} + \ion{H}{i} $\rightarrow$ \ion{Fe}{ii} + ${\rm H}^-$ using similar symbols. The calculations were made with $T = 5830$~K, log~$N_{\rm e}$(cm$^{-3}$) = 13, and log~$N_{\rm H}$(cm$^{-3}$) = 16.9.
   }
 \label{fig:rates1}
 \end{center}
% \end{minipage} 
 \end{figure*}

\subsection{Updated model atom} \label{Sect:model}

This study updates the NLTE method developed by \citet[][hereafter, Paper~I]{mash_fe}. First, we briefly describe the atomic data  taken from Paper~I.
The model atom includes 239 levels of \ion{Fe}{i}, 89 levels of \ion{Fe}{ii}, and the ground state of \ion{Fe}{iii}. The multiplet fine structure is neglected. The model levels of \ion{Fe}{i} represent the measured levels, belonging to 233 terms, and the predicted high-excitation (\Eexc\ $>$ 7.1~eV) levels, out of which the six super-levels were built up. The \ion{Fe}{ii} levels belong to 89 terms with \Eexc\ up to 10~eV. 
We use transition probabilities from the \citet[][\ion{Fe}{i}]{1994ApJS...94..221N} compilation and calculations of \citet{Kurucz2009} and \citet{1992RMxAA..23...45K} for \ion{Fe}{i} and \ion{Fe}{ii}, respectively. 
%All levels in the model atom are coupled via collisional excitation and ionisation by electrons. 
For the transitions between the \ion{Fe}{ii} terms up to \eu{z}{4}{D}{\circ}{}, electron-impact excitation data are taken from the close-coupling calculations of \citet{1995A&A...293..953Z,1996A&AS..115..551B}, and \citet{1998ApJ...492..650B}.
   Electron-impact ionisation cross-sections are calculated from the 
%\citet{1962amp..conf..375S} 
classical path approximation, 
%{\bf using formula 9.60 from \citet{2014tsa..book.....H}. }.
with a mean Gaunt factor of $\overline{g}$ = 0.1 for \ion{Fe}{i} and to 0.2 for \ion{Fe}{ii}, as recommended by \citet{1962amp..conf..375S}. 

 The modifications concern computations of photoionisation and collisional rates.
 For 116 levels of \ion{Fe}{i} their photoionisation cross-sections are taken from the $R$-matrix calculations of 
%\citet{1997A&AS..122..167B} and 
\citet{2017A&A...606A.127B}.
For the remaining levels of \ion{Fe}{i} and for all the \ion{Fe}{ii} levels, we adopt the hydrogenic approximation, where principal quantum number of the level is replaced with the corresponding effective principal quantum number. 

Compared with Paper~I, a treatment of electron-impact excitation is updated for 1031 transitions of \ion{Fe}{i}, with employing the collision strengths from R-matrix calculations of \citet{2017A&A...606A.127B}. 
%For all the transitions of \ion{Fe}{i} and \ion{Fe}{ii}, with no accurate collisional data available, we rely on theoretical approximations, as in Paper~I. 

A novelty of this research is that we take into account inelastic processes in collisions with \ion{H}{i} atoms for not only \ion{Fe}{i}, but also \ion{Fe}{ii}, using
%the ion-pair production from the energy levels of \ion{Fe}{i} and mutual neutralisation (charge-exchange reactions),
%\ion{Fe}{i}($nl$) + \ion{H}{i} $\leftrightarrow$ \ion{Fe}{ii}(\eu{a}{6}{D}{}{}, \eu{a}{4}{F}{}{}) + H$^-$
%\noindent
%and \ion{H}{i} impact excitation and de-excitation processes in \ion{Fe}{i} and \ion{Fe}{ii}, with 
the rate coefficients from quantum-mechanical calculations. 

For 
%\ion{Fe}{i} + \ion{H}{i}, 
the ion-pair production from the energy levels of \ion{Fe}{i} and mutual neutralisation (charge-exchange reactions),

\ion{Fe}{i}($nl$) + \ion{H}{i} $\leftrightarrow$ \ion{Fe}{ii}(\eu{a}{6}{D}{}{}, \eu{a}{4}{F}{}{}) + H$^-$

\noindent
and \ion{H}{i} impact excitation and de-excitation processes in \ion{Fe}{i}, 
the required data were taken from two different studies. \citet{2018A&A...612A..90B} used a method based on an asymptotic two-electron linear combination of atomic orbitals model of ionic-covalent interactions in the neutral atom-hydrogen-atom system, together with the multi-channel Landau-Zener model, and performed calculations including 166 covalent states of \ion{Fe}{i} and 25 ionic states. Hereafter, we refer to these data as B18. In the SE calculations with the B18 data, we take into account 58 charge-exchange reactions, where the \ion{Fe}{ii} ionic state is either the ground \eu{a}{6}{D}{}{} or the first excited \eu{a}{4}{F}{}{} state. At given temperature, rate coefficients of the remaining charge-exchange reactions are several orders of magnitude smaller, such that their influence on the SE of iron can be neglected.
The rate coefficients of \citet[][hereafter, YBK18]{2018CP....515..369Y} come from the quantum simplified model approach that was applied to low-energy collisions of iron atoms and cations with hydrogen atoms and anions. In total, 97 low-lying covalent Fe + H states and two ionic \ion{Fe}{ii} + \ion{H}{i} molecular states were treated. 

It is important to compare the results of B18 and YBK18. 
Although the approaches used are different (YBK18 is based on the quantum asymptotic semi-empirical approach), the test calculations \citep{2018MNRAS.473.3810Y} have shown that the models perform equally well, on average. 
%leading to atomic data roughly of the same accuracy. 
  This does not mean that the results are identical. This means that atomic data calculated by these two methods are roughly of the same accuracy, but using different methods leads to some scatter in the H-collision rate coefficients.
The main difference of the two approaches applied in B18 and YBK18 is treating transitions that involve changing of the Fe$^+$ core. Both approaches are based on the ionic-covalent interactions. Within the two-electron linear combination of atomic orbital framework used in B18, the off-diagonal matrix elements are equal to zero between covalent and ionic states that have different Fe$^+$ cores, and, therefore, rate coefficients for corresponding processes are equal to zero as well. The simplified model used in YBK18 is free from this limitation since it is based on the semi-empirical formula for off-diagonal Hamiltonian matrix elements whatever the number of electrons is. According to the general rule, the off-diagonal matrix elements are non-zero, if the ionic and covalent states have the same molecular symmetry. The semi-empirical formula does not provide estimates for matrix elements with different cores, but does for single-electron transitions. So, one can estimate core-changed matrix elements by means of the semi-empirical formula treating such data as upper-limit estimates, these are the YBK18 data, while the results of B18 with zero core-changed matrix elements should be treated as lower-limit estimates. In addition, both methods do not take short-range non-adiabatic regions into account. It is known that accounting for short-range regions may increase rate coefficients by up to several orders of magnitude. Treating upper-limit data compensates somehow not accounting for the short-range regions, and this is another reason to use the upper-limit data. Thus, the results of both methods can be considered as lower-limit and upper-limit estimates for the inelastic rate coefficients.

Figure~\ref{fig:rates1} (left panel) displays the \ion{Fe}{i} excitation rates depending on the transition energy, $E_{lu}$, for electron impact and \ion{H}{i} impact with the rate coefficients from two sources, YBK18 and B18. 
%For comparison, we show also the Drawinian rates scaled by \kH\ = 0.5. Here, 
The data correspond to a kinetic temperature of $T = 5830$~K and an \ion{H}{i} number density of log~$N_{\rm H}$(cm$^{-3}$) = 16.9 that are characteristic of the line-formation layers (log~$\tau_{5000} = -0.54$) in the model with $\Teff$ / $\logg$ / [Fe/H] = 6350~K / 4.09 / $-2.15$, which represents the atmosphere of one of our sample stars, HD~84937. In general, collisional rates grow towards smaller $E_{lu}$, although, in each collisional recipe, log~C can differ by 2-5 dex for the transitions of close energy. Compared with electron impacts, collisions with \ion{H}{i} are more efficient in exciting the $E_{lu} \lesssim$ 2~eV  transitions, independent of using either B18 or YBK18 rate coefficients.

The right panel of Fig.~\ref{fig:rates1} shows the ion-pair production and the electron-impact ionisation rates depending on the level ionisation energy.
%, $\chi_l$. It can be seen that 
The charge-exchange reactions are much more efficient than the collisional ionisation and their inverse processes in coupling \ion{Fe}{i} to \ion{Fe}{ii}.
%ion-pair production rates are substantially higher than  for atomic levels with $\chi_l \simeq$ 1.3 to 3.5~eV (.

\begin{figure}
% \begin{minipage}{190mm}
 \begin{center}
\resizebox{90mm}{!}{\includegraphics{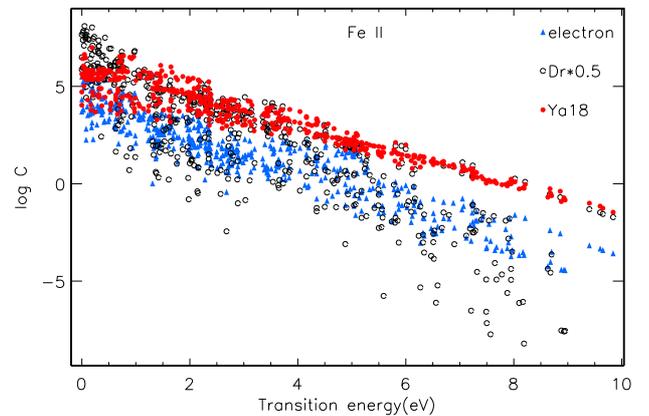}}
  \caption{\ion{Fe}{ii} excitation rates (in s$^{-1}$), log~C, for electron impact (triangles) compared with the rates for \ion{H}{i} collisions from calculations of YBK19 (filled circles) and compared with the scaled (\kH\ = 0.5) Drawinian rates (open circles). The calculations were made with $T = 5830$~K, log~$N_{\rm e}$(cm$^{-3}$) = 13, and log~$N_{\rm H}$(cm$^{-3}$) = 16.9.
   }
 \label{fig:rates2}
 \end{center}
% \end{minipage} 
 \end{figure}

The quantum simplified model including 117 covalent and the ground ionic states was applied by \citet[][hereafter, YBK19]{2019MNRAS.483.5105Y} to calculate the rate coefficients for inelastic processes in low-energy \ion{Fe}{ii} + \ion{H}{i} collisions. Using these data, we take into account the \ion{H}{i} impact excitation for 528 transitions of \ion{Fe}{ii} in our model atom. We do not take into account the charge-exchange reactions, \ion{Fe}{ii} + \ion{H}{i} $\rightleftarrows$ \ion{Fe}{iii} + ${\rm H}^-$,
 because of very low number density of \ion{Fe}{iii}. For example, N(\ion{Fe}{iii})/N(Fe) $< 10^{-3}$ everywhere in the model 6350/4.09/$-2.15$. 
 Figure~\ref{fig:rates2} shows the electron impact and \ion{H}{i} impact excitation rates for transitions of \ion{Fe}{ii}. At given transition energy, collisions with \ion{H}{i} are more numerous than electronic collisions, and only at $E_{lu} <$ 0.5~eV the two types of collisional rates approach each other. Thus, \ion{Fe}{ii} + \ion{H}{i} collisions are expected to serve as efficient source of thermalisation in the atmospheres of cool stars. We note that the quantum mechanical collisional rates lie at the upper boundary of the Drawinian rate set.

%The simplified model approach is applied to low-energy collisions of iron atoms and cations with hydrogen atoms and anions.Inelastic collisional processes for all transitions involving 97 low-lying covalent Fe + H states and two ionic Fe+ + H- molecular states are treated,

For \ion{H}{i} impact excitation of the \ion{Fe}{i} transitions missing in YBK18 and B18, we employ the rate coefficients computed by \citet{2017ascl.soft01005B} based on free electron model of \citet {1991JPhB...24L.127K}. The data are accessible at {\tt https://github.com/barklem/kaulakys/}.

We emphasise that collisions with \ion{H}{i} are neglected for those \ion{Fe}{i} and \ion{Fe}{ii} transitions, for which none of the cited sources provides rate coefficients.
%In Section~\ref{sec:stars}, we inspect the influence of employing both YBK18 and B18 rate coefficients for \ion{Fe}{i} and the data of YBK19 for \ion{Fe}{ii} on the iron abundance determinations. 

\subsection{Statistical equilibrium of \ion{Fe}{i} - \ion{Fe}{ii}}\label{Sect:effect}

 \begin{figure}
% \begin{minipage}{190mm}
 \begin{center}
 \resizebox{90mm}{!}{\includegraphics{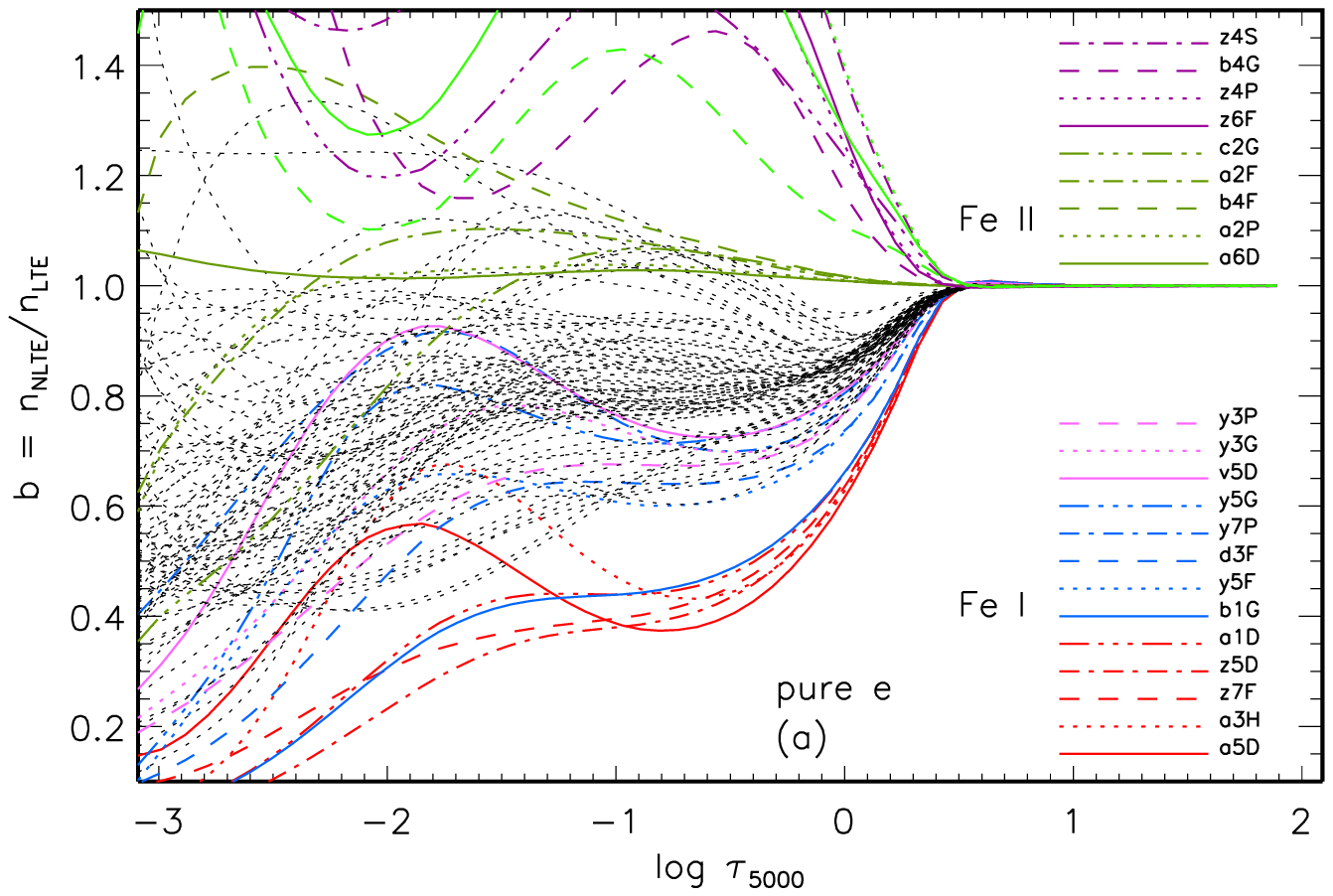}}
 \resizebox{90mm}{!}{\includegraphics{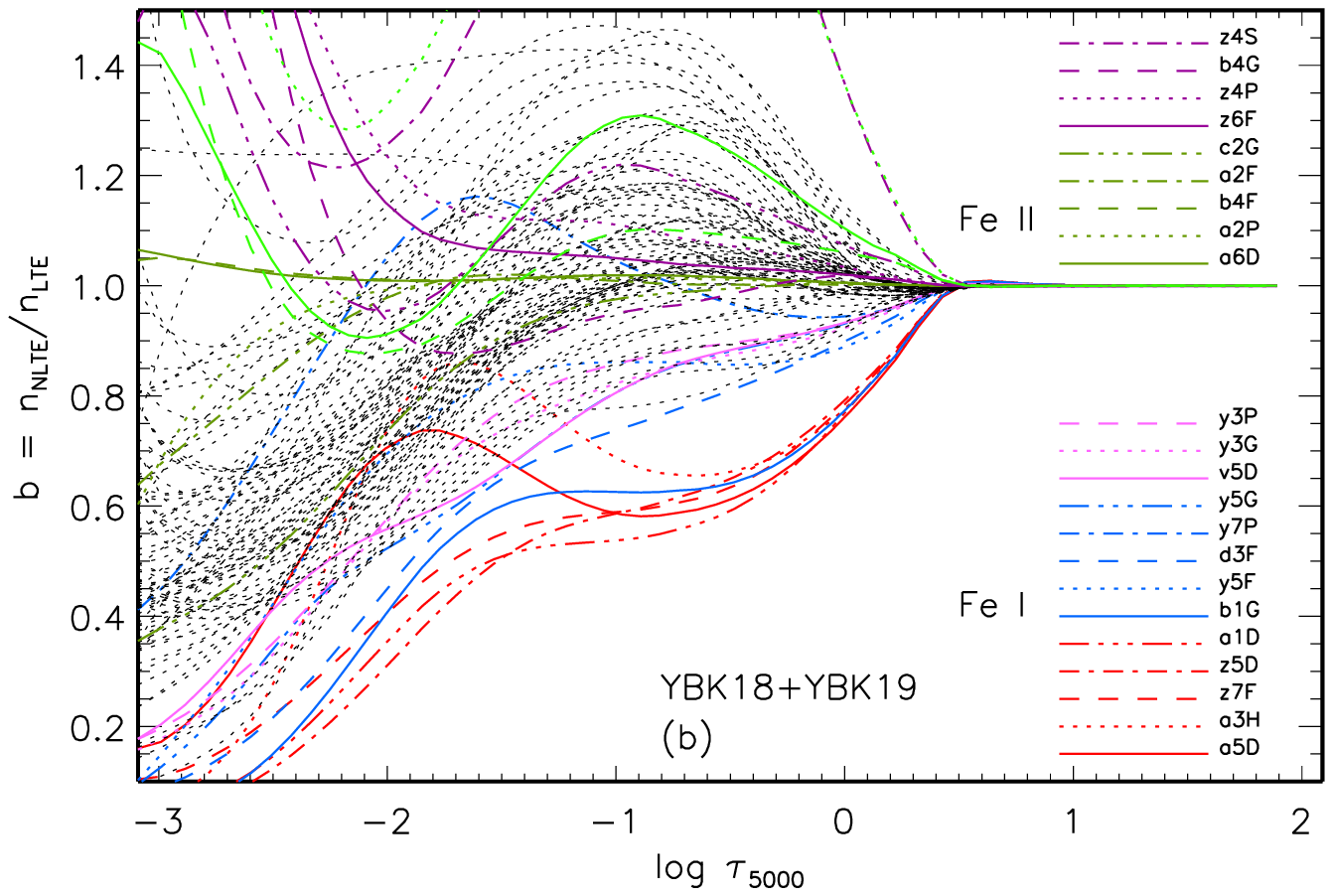}}
 \resizebox{90mm}{!}{\includegraphics{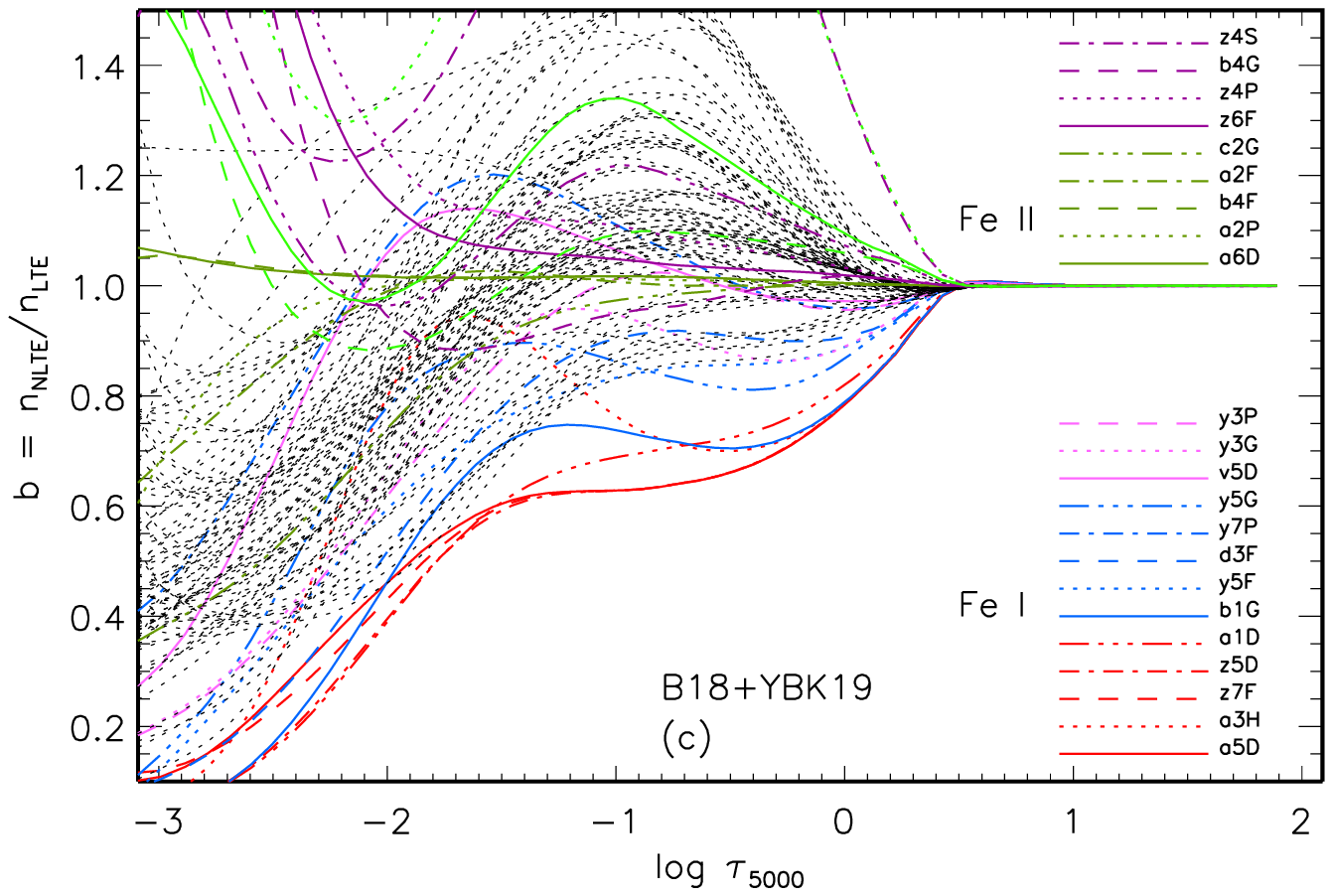}}
\caption{Departure coefficients, b, for the levels of \ion{Fe}{i} and \ion{Fe}{ii} as a function of $\log \tau_{5000}$ in the model atmosphere 4600/1.40/$-2.60$ from calculations using different treatment of \ion{H}{i} collisions: pure electronic collisions (top panel); YBK18 + YBK19 (middle panel); B18 + YBK19 (bottom panel). Every fifth of the first 60 levels (\Eexc\ $\le$ 5.73~eV) of \ion{Fe}{i} is shown. They are quoted in the bottom right part of each panel. All levels of \ion{Fe}{i} above \Eexc\ = 7~eV are plotted by black dotted curves. For \ion{Fe}{ii}  we show every fifth of the first 60 levels (\Eexc\ $\le$ 8.16~eV). They are quoted in the top right part of each panel. The letters z, y, x, ... are used to denote the odd parity terms and a, b, c, ... for the even parity terms.} \label{Fig:bf_fe12}
 \end{center}
\end{figure}

We choose the VMP model atmosphere, 4600/1.40/$-2.60$, to investigate the influence of inelastic collisions with \ion{H}{i} and their different treatment on the SE of \ion{Fe}{i}-\ion{Fe}{ii}. 
The calculations were performed for three different line-formation scenarios based on pure electronic collisions (a), including \ion{H}{i} collisions for \ion{Fe}{i} with the rate coefficients from YBK18 (b) and B18 (c). In both (b) and (c) cases, \ion{Fe}{ii} + \ion{H}{i} collisions are accounted for using the data of YBK19. 
Figure\,\ref{Fig:bf_fe12} displays the departure coefficients, ${\rm b = n_{NLTE}/n_{LTE}}$, for selected levels of \ion{Fe}{i} and \ion{Fe}{ii}. Here,
${\rm n_{NLTE}}$ and ${\rm n_{LTE}}$ are the statistical equilibrium and thermal (Saha-Boltzmann) number densities, respectively. 

As shown in the earlier NLTE studies \citep[see][and references therein]{mash_fe}, the main NLTE mechanism for \ion{Fe}{i} in the stellar parameter range, with which we are concerned here, is the ultra-violet (UV) over-ionisation caused by superthermal radiation of a non-local origin below the thresholds of the low excitation levels. It can be seen in Fig.\,\ref{Fig:bf_fe12}
that the \ion{Fe}{i} levels below \eu{d}{3}{F}{}{} (\Eexc\ $\simeq$ 4.5~eV) are underpopulated at $\log \tau_{5000} < 0.4$, independent of the treatment of collisional rates. In case of pure electronic collisions, even the highest levels of \ion{Fe}{i} are strongly decoupled from the ground state of \ion{Fe}{ii}. As expected, the departures from LTE are smaller when collisions with \ion{H}{i} are included. For example, at $\log \tau_{5000} = -0.52$, the departure coefficients of \eu{a}{3}{F}{}{} and \eu{z}{5}{F}{\circ}{}, which are the lower levels of the transitions, where the \ion{Fe}{i} 5216 and 5586\,\AA\ lines arise, amount to b = 0.36 and 0.40 in case of pure electronic collisions, while they increase in the (b) and (c) scenarios: b(\eu{a}{3}{F}{}{}) = 0.66 and 0.60 and b(\eu{z}{5}{F}{\circ}{}) = 0.71 and 0.69, respectively. 

All the levels of \ion{Fe}{ii} below \eu{c}{2}{D}{}{} (\Eexc\ = 4.73~eV) have a common parity, and, independent of the treatment of collisional rates, they are closely coupled to the \ion{Fe}{ii} ground state throughout the atmosphere except the outermost layers. The odd-parity levels with \Eexc\ $\ge$ 4.8~eV are affected by the pumped UV transitions from the ground and low-excitation levels of \ion{Fe}{ii}. In case of pure electronic collisions, this results in large overpopulation of all the levels above \Eexc\ = 4.8~eV. Including collisions with \ion{H}{i} reduces substantially the departures from LTE, in particular, for the levels below \eu{b}{4}{G}{}{} (\Eexc\ = 6.7~eV). 

\section{Iron abundances and \ion{Fe}{i} / \ion{Fe}{ii} ionisation equilibrium of reference stars}\label{sec:stars}

In this section, we test the updated model atom with the \ion{Fe}{i} /\ion{Fe}{ii} ionisation equilibrium of the VMP stars with well determined atmospheric parameters. We select three Galactic halo  stars, namely, HD~84937, HD~122563, and HD~140283, and a sample of 38 VMP stars in the dwarf spheroidal galaxies (dSphs) from \citet[][35 stars]{dsph_parameters} and \citet[][3 stars]{Pakhomov_boo1}.

\subsection{Atmospheric parameters}

Our analysis is based on photometric effective temperatures and distance based surface gravities.
For a VMP giant HD~122563, coupling its angular diameter with photometry yields effective temperatures, which are fairly consistent in \citet{2012A&A...545A..17C}: $\Teff$ = 4598$\pm$41~K and \citet{2018MNRAS.475L..81K}: $\Teff$ = 4636$\pm$37~K. 
 Using a Gaia based distance of $d$ = 288~pc from \citet{2018AJ....156...58B} and assuming the star's mass $M = 0.8 M_\odot$, we calculate $\logg_d$ = 1.42$\pm$0.02. This value agrees well with $\logg$ = 1.39$\pm$0.01 based on the detections of sun-like oscillations \citep{2019arXiv190202657C}. We adopt $\Teff$ = 4600~K and $\logg$ = 1.40 in our calculations. Based on our analysis of HD~122563 in Paper~I, we use the model atmosphere with [Fe/H] = $-2.6$. Both in NLTE and LTE, the slope of the $\eps{\rm FeI}$ -- log W$_{\rm obs}/\lambda$ plot is largely removed with a microturbulent velocity of $\xi_t$ = 1.6~\kms.  
 %From a variation in the \ion{Fe}{i} based abundance caused by a variation $\xi_t$, 
 An accuracy of the $\xi_t$ determination is estimated as 0.2~\kms. We note that our $\xi_t$ value is consistent with the microturbulent velocity derived by \citet{Bergemann_fe_nlte} and agrees within the error bars with that obtained by \citet{2016MNRAS.463.1518A} in their 1D-NLTE analysis. Hereafter, $\eps{\rm FeII}$ and $\eps{\rm FeI}$ are the \ion{Fe}{ii} and the \ion{Fe}{i} based abundances, respectively. We use the scale where $\eps{H}$ = 12.
  % with interferometric measurements of  and the trigonometric parallax known from the Gaia DR2 \citep{2018A&A...616A...1G}. 

For HD~84937 and HD~140283, we take their $\Teff$ and $\logg$ from a careful analysis of \citet[][hereafter, Paper~II]{2015ApJ...808..148S}. For HD~140283, the adopted atmospheric parameters are supported by recent measurements: \citet{2018MNRAS.475L..81K} determine $\Teff$ = 5787$\pm$48~K and a use of the Gaia DR2 parallax \citep{2018A&A...616A...1G} yields $\logg$ = 3.66$\pm$0.03. In this study, we revise stellar metallicities, [Fe/H], using the \ion{Fe}{ii} lines in the visible spectral range and the solar abundance $\eps{\odot,FeII}$ = 7.54 (Paper~II), which is based on $gf$-values of \citet{RU}. The microturbulent velocities were revised from a requirement that \ion{Fe}{i} lines of different strength yield consistent absolute abundances. An accuracy of the $\xi_t$ determination is estimated as 0.2~\kms. For HD~84937, our $\xi_t$ value agrees well with that obtained by \citet{2016MNRAS.463.1518A} in their 1D-NLTE analysis, while it is larger than that of \citet{Bergemann_fe_nlte}, by 0.3~\kms. In contrast, our $\xi_t$ value for HD~140283 is consistent with that of \citet{Bergemann_fe_nlte}. 
  
For the dSph stars, we adopt their $\Teff$, $\logg$, [Fe/H], and $\xi_t$, as determined by \citet{dsph_parameters} and \citet{Pakhomov_boo1}. We do not revise microturbulent velocities because applying new collisional data affects only a little the slopes of the $\eps{\rm FeI}$ -- log W$_{\rm obs}/\lambda$ trends compared with the corresponding values in our previous studies. 

\begin{table*} %[htbp]
\caption{Atmospheric parameters and iron abundances, $\eps{Fe}$, of the investigated stars. \label{tab_param}}
 \centering
 \begin{tabular}{lcccclcccccc}
\hline\hline \noalign{\smallskip}
Star & $\Teff$ & $\logg$ & [Fe/H] & $\xi_t$ & & \multicolumn{2}{c}{LTE} & & \multicolumn{3}{c}{NLTE} \\
\cline{7-8} \cline{10-12}
     &  K      & CGS     &        & \kms    & & \ion{Fe}{i} & \ion{Fe}{ii} & & \ion{Fe}{i} (YBK18)  & \ion{Fe}{i} (B18) & \ion{Fe}{ii} \\
\hline
\multicolumn{1}{c}{(1)} & (2) & (3) & (4) & (5) & & (6) & (7) & & (8) & (9) & (10) \\
\noalign{\smallskip} \hline \noalign{\smallskip}
% Sun     & 5777  & 4.44  & & 0.9 & vis & 7.53(0.09)$^1$ & 7.56(0.05) & & 7.55(0.10) & 7.56(0.09) & 7.56(0.05) \\
 HD~84937   & 6350 & 4.09 & $-2.18$ & 1.7 & vis & 5.25(0.05) & 5.36(0.08) & & 5.47(0.08) & 5.49(0.08) & 5.36(0.08) \\ 
 HD~84937   &      &      &         &     & UV  & 5.22(0.12) & 5.30(0.10) & & 5.39(0.12) & 5.42(0.12) & 5.30(0.10) \\   
 HD~84937   &      &      &         &\multicolumn{2}{r}{vis+UV} & 5.23(0.11) & 5.31(0.10) & & 5.41(0.12) & 5.44(0.11) & 5.31(0.10) \\   
 HD~122563  & 4600 & 1.40 & $-2.55$ & 1.6 & vis & 4.73(0.15) & 4.98(0.07) & & 4.91(0.11) & 4.89(0.14) & 4.98(0.08) \\            
 HD~140283  & 5780 & 3.70 & $-2.43$ & 1.3 & vis & 4.95(0.09) & 5.11(0.07) & & 5.21(0.09) & 5.23(0.09) & 5.11(0.07) \\
 HD~140283  &      &      &         &     & UV  & 4.98(0.12) & 5.08(0.10) & & 5.14(0.11) & 5.18(0.11) & 5.08(0.10) \\   
 HD~140283  &      &      &         &\multicolumn{2}{r}{vis+UV} & 4.97(0.11) & 5.08(0.10) & & 5.17(0.11) & 5.20(0.11) & 5.08(0.10) \\   
%\hline
\noalign{\smallskip}\hline \noalign{\smallskip}
\multicolumn{12}{l}{{\bf Notes.} $^1$ The numbers in parentheses are the dispersions in the single line measurements around the mean. } \\
\end{tabular}
\end{table*}

\begin{figure*}
% \begin{center}
%\hspace{-0.1\linewidth}
  \resizebox{90mm}{!}{\includegraphics{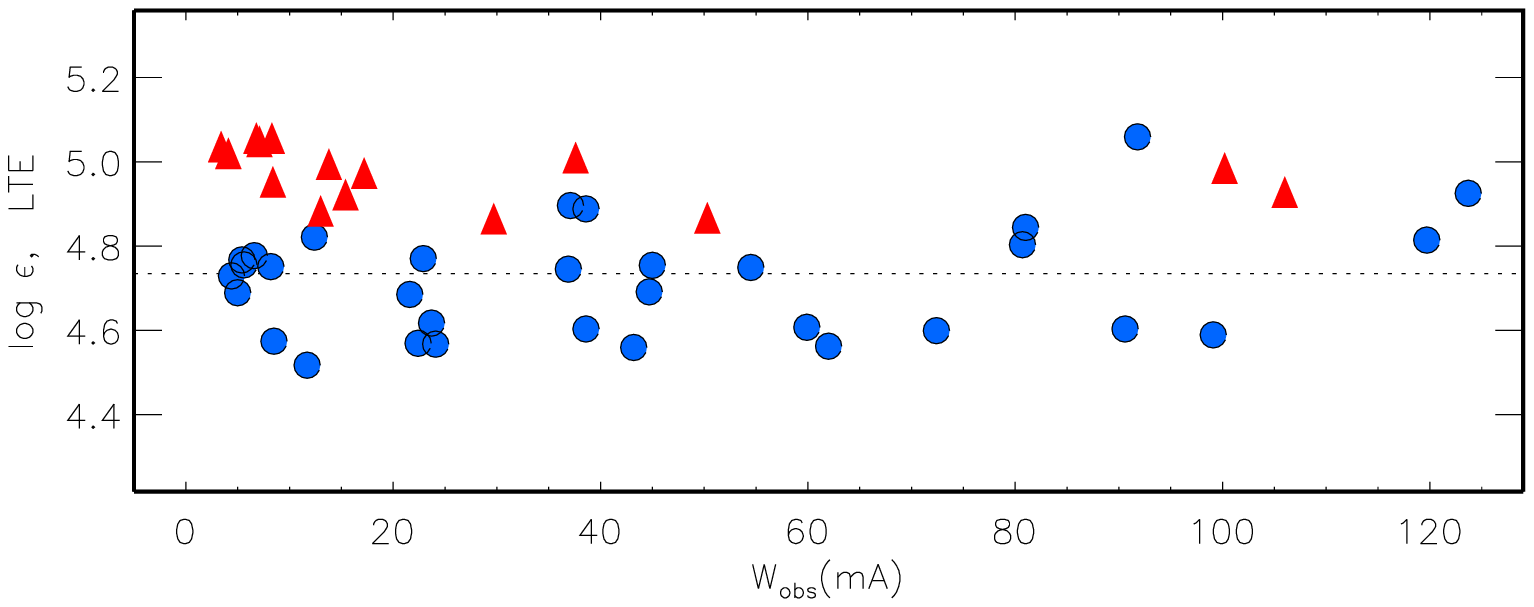}}
  \resizebox{90mm}{!}{\includegraphics{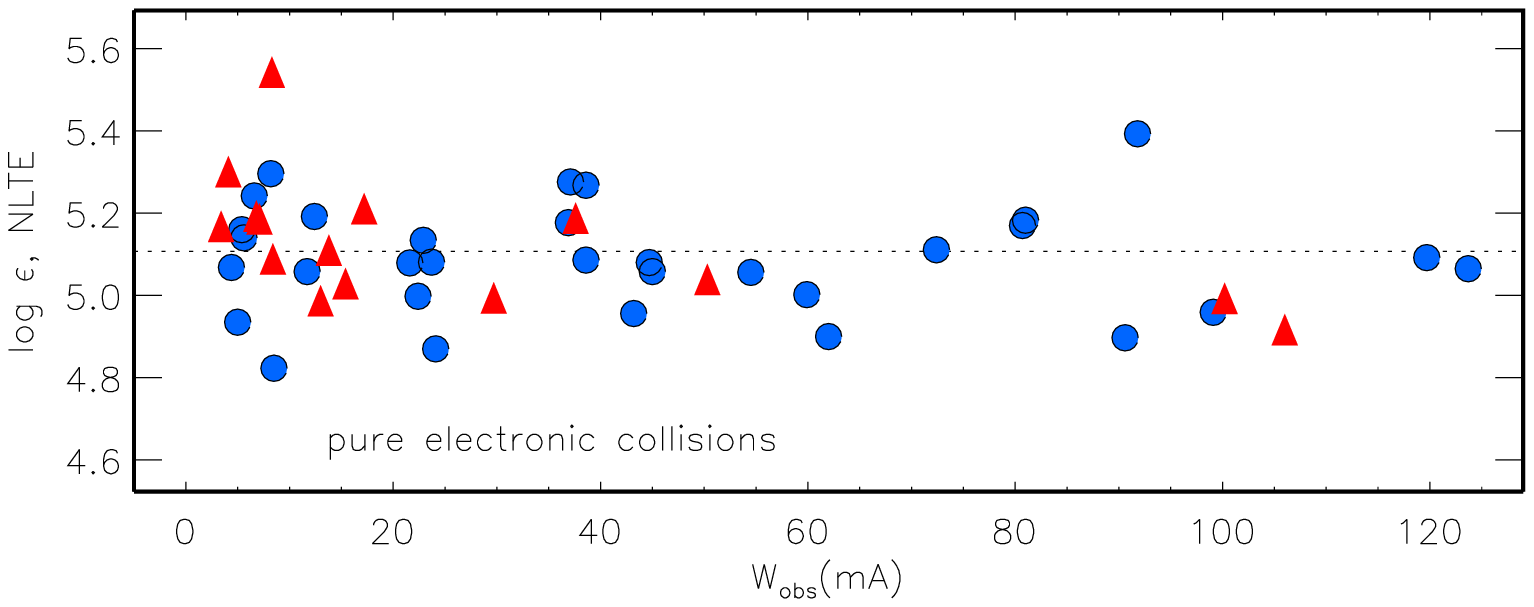}}

%  \vspace{-13mm}  
  \resizebox{90mm}{!}{\includegraphics{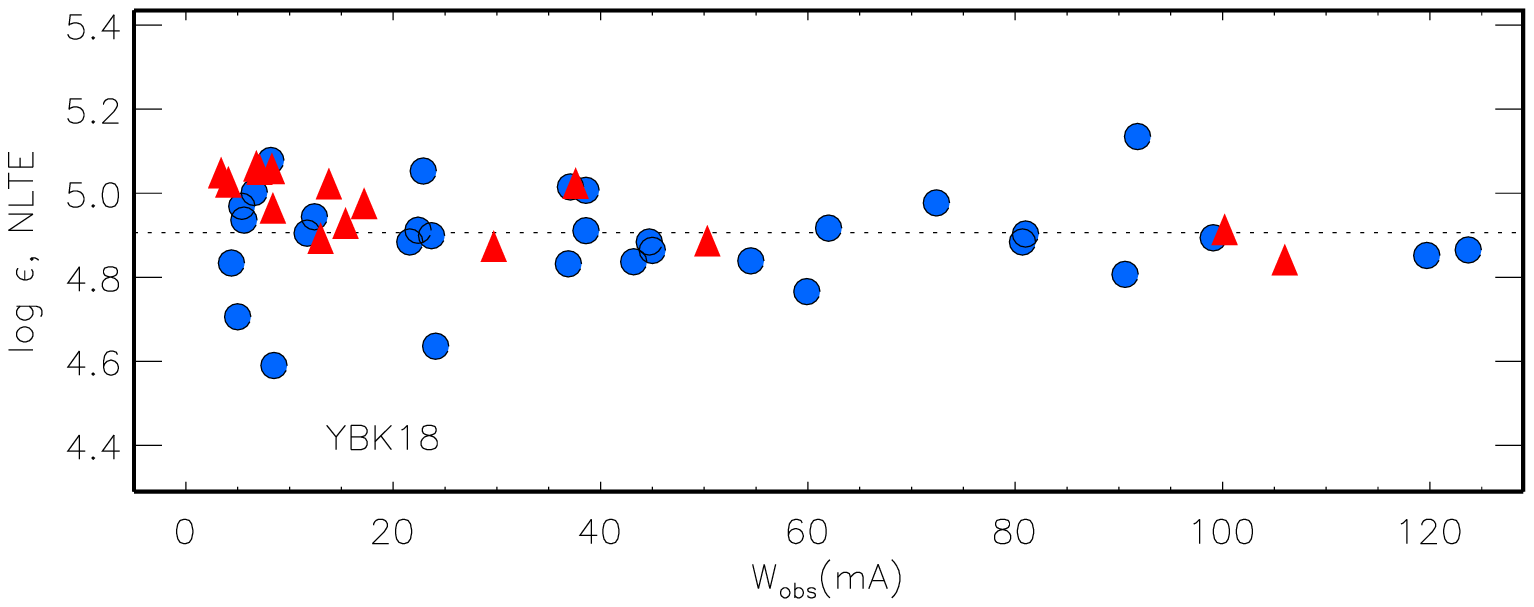}}
  \resizebox{90mm}{!}{\includegraphics{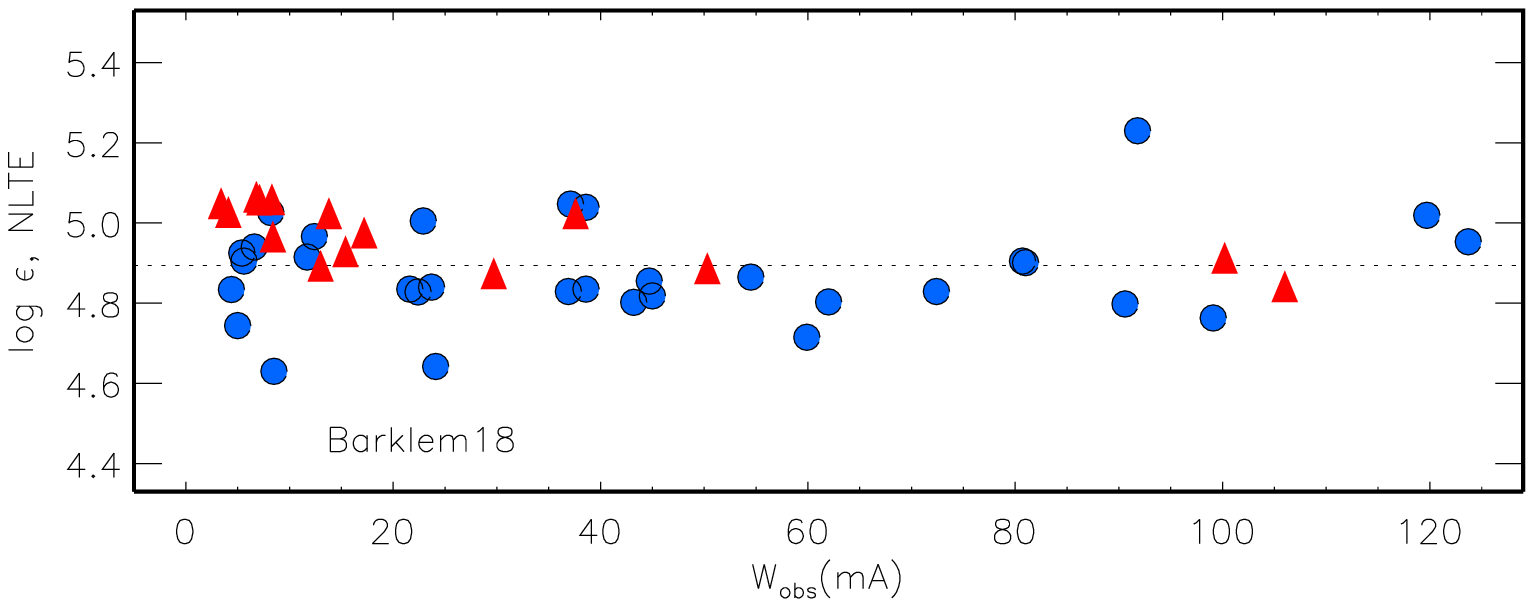}}
 \caption{LTE (top left panel) and NLTE (other three panels) abundances from lines of \ion{Fe}{i} (circles) and \ion{Fe}{ii} (triangles) in HD~122563 as a function of observed equivalent width W$_{\rm obs}$. The top right panel corresponds to the NLTE calculations with pure electronic collisions. In the bottom row, the left and right panels correspond to the YBK18+YBK19 and B18+YBK19 scenarios, respectively.
 % \ion{Fe}{i} + \ion{H}{i} collisions are treated following YBK18 in the left panel and following \citet{2018A&A...612A..90B} in the . The \ion{Fe}{ii} + \ion{H}{i} collisions are treated following YBK19. 
 In each panel, the dotted line indicates the mean abundance derived from the \ion{Fe}{i} lines.} 
 \label{fig:hd122563}
% \end{center}
 \end{figure*}

\subsection{Methods of abundance determinations}

For the halo benchmark stars, both LTE and NLTE abundances are determined from line profile fitting.
The synthetic spectra are computed with the {\sc Synth}V\_NLTE code \citep{2019ASPC} that implements the pre-computed departure coefficients from the {\sc DETAIL} code. The best fit to the observed spectrum is obtained automatically with using 
the {\sc IDL binmag} code by O. Kochukhov\footnote{http://www.astro.uu.se/$\sim$oleg/download.html}.  The line list and atomic data for the synthetic spectra calculations are taken from the VALD database \citep{2015PhyS...90e4005R}.

For the dSph stars, the NLTE abundances from individual lines are calculated by applying the NLTE abundance corrections computed in this study to the LTE abundances derived by \citet{dsph_parameters} and \citet{Pakhomov_boo1}. 

\subsection{Galactic halo benchmark stars}\label{sec:halo}

The atmospheric parameters and the average abundances for the two ionisation stages are presented in Table~\ref{tab_param}. Abundances from individual lines are given in Table~\ref{tab:lines} (Table~A.1, online material).

 \underline{HD~122563}.
We use 32 lines of \ion{Fe}{i} and 15 lines of \ion{Fe}{ii} in a high-quality spectrum from the ESO UVESPOP survey \citep{2003Msngr.114...10B}. For \ion{Fe}{i}, the line atomic data are taken from Paper~I (their Table~5).  
%For \ion{Fe}{i}, $gf$-values are mostly from \citet{1991JOSAB...8.1185O}. 
The exceptions are \ion{Fe}{i} 5242, 5662, and 5638\,\AA, for which we adopt the most recent $gf$-values from \citet{2017ApJ...848..125B,2014ApJS..215...23D}, and \citet{2014MNRAS.441.3127R}, respectively. We note that, for each of these three lines, their $gf$-values were upward revised, by 0.09 to 0.15 dex, and this leads to larger line-to-line scatter in the W$_{\rm obs} <$ 12\,m\AA\ equivalent width range (Fig.\,\ref{fig:hd122563}) compared with using the older data. For \ion{Fe}{ii}, we apply $gf$-values of \citet{RU}. The exceptions are \ion{Fe}{ii} 4923 and 5018\,\AA, for which $gf$-values were obtained by averaging the data from the four sources with both lines available, namely, \citet{Bridges_fe2,1983A&AS...52...37M}, \citet[][hereafter, RU98]{RU}, and \citet[][hereafter, MB09]{MB09}. 

 In contrast to Paper~I, we determine here absolute, but not differential abundances. They are derived in five different line-formation scenarios, namely: LTE, NLTE with pure electronic collisions, NLTE with including \ion{H}{i} collisions and using the rate coefficients from YBK19 for \ion{Fe}{ii} and YBK18 or B18 for \ion{Fe}{i}, and NLTE based on the Drawinian rates. 
Abundances from individual lines are shown in Fig.~\ref{fig:hd122563} for the first four cases.
%, while   Table~\ref{tab_param} presents the mean 
%LTE and NLTE (scenarios YBK18 and B18) 
%abundances for each ionisation stage. 
%As in Paper~I, we applied a line-by-line differential NLTE and LTE approach, in the sense that stellar line abundances were compared with individual abundances of their solar counterparts. 
We comment on the results obtained in different line-formation scenarios. 
 
{\it LTE:} an abundance difference of $-0.25$~dex is found between \ion{Fe}{i} and \ion{Fe}{ii}. 

\begin{figure}
 \begin{center}
  \resizebox{90mm}{!}{\includegraphics{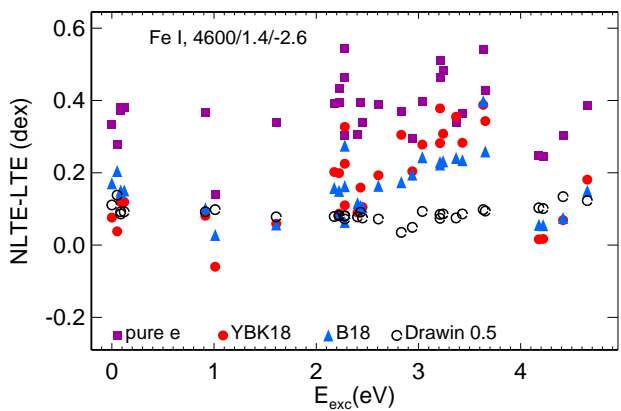}}
\caption{NLTE abundance corrections for the lines of \ion{Fe}{i} in the 4600/1.40/$-2.60$ model from calculations that use pure electronic collisions (squares) and include collisions with \ion{H}{i} according to YBK18 (filled circles) and B18 (triangles). 
For comparison, the NLTE corrections were computed with the scaled Drawinian rates (\kH\ = 0.5, open circles).}  \label{Fig:dnlte}
 \end{center}
\end{figure}

{\it Pure electronic collisions.} The NLTE effects for lines of both \ion{Fe}{i} and \ion{Fe}{ii} are substantially larger compared with that in the other NLTE scenarios (Fig.~\ref{Fig:dnlte}, only \ion{Fe}{i}). For lines of \ion{Fe}{i} the NLTE abundance corrections, $\Delta_{\rm NLTE}  = \eps{NLTE} - \eps{LTE}$, are positive, and  
 the obtained mean abundance, $\eps{\rm FeI}$ = 5.11$\pm0.14$, is 0.38~dex higher than the LTE value. Here, the sample standard deviation, $\sigma = \sqrt{\Sigma(\overline{x}-x_i)^2 / (N_l-1)}$, determines the dispersion in the
single line measurements around the mean for given ionisation stage and $N_l$ is the number of measured lines. As discussed in Sect.~\ref{Sect:effect}, the \ion{Fe}{ii} levels above \Eexc\ = 4.8~eV have large overpopulations in case of pure electronic collisions. This results in positive NLTE corrections of $\Delta_{\rm NLTE}$ = 0.09 to 0.47~dex for lines of \ion{Fe}{ii}. The exceptions are \ion{Fe}{ii} 4923 and 5018~\AA, for which $\Delta_{\rm NLTE}$ is slightly negative, of $-0.01$ and $-0.03$~dex, respectively. The mean NLTE abundance, $\eps{\rm FeII}$ = $5.16\pm0.21$, is 0.18~dex higher than the LTE one, and we note large line-to-line scatter, resulting in $\sigma$, which is more than twice larger than in LTE.

%, independent of treatment of collisional rates. However, They are substantially larger in case of pure electronic collisions, such that
%Figure~\ref{Fig:dnlte} displays the NLTE abundance corrections for lines of \ion{Fe}{i} in each of the NLTE scenarios.

{\it Electronic + hydrogenic collisions.}
Independent of using either YBK18 or B18 data for \ion{Fe}{i} + \ion{H}{i} collisions, abundances from the two ionisation stages of iron are found to agree within the error bars.  
Although 
%when collisions with \ion{H}{i} are included using recipes of either YBK18 or B18 for \ion{Fe}{i} and YBK19 for \ion{Fe}{ii}. In both cases, YBK18 and B18, 
the NLTE abundance corrections for individual lines of \ion{Fe}{i} can differ in these two scenarios, by up to 0.1~dex (Fig.~\ref{Fig:dnlte}). The exceptions are the \ion{Fe}{i} 4427, 4920, and 5324\,\AA\ lines, for which the difference in $\Delta_{\rm NLTE}$ between B18 and  YBK18 amounts to 0.17, 0.13, and 0.15~dex, respectively.
%(\ion{Fe}{i} 5324~\AA). 

%reveal a similar behavior as a function of\Eexc, although a spread of $\Delta_{\rm NLTE}$ is larger in the YBK18 case. 
Including collisions with \ion{H}{i} reduces substantially the NLTE effects for lines of \ion{Fe}{ii}, and $\Delta_{\rm NLTE}$s are only slightly positive (0.00 to 0.03~dex).
%compared with that for pure electronic collisions.  
%the NLTE abundance corrections are much smaller than that for \ion{Fe}{i} and that for \ion{Fe}{ii} in the calculations with 
%for most investigated lines of \ion{Fe}{ii} .
% This is due to enhanced populations of the upper levels relatively to the populations of the lower levels in the line-formation layers (Fig.~\ref{Fig:bf_fe12}). 
The exceptions are the strongest \ion{Fe}{ii} 4923 and 5018~\AA\ lines. 
Their cores form in the atmospheric layers, where the departure coefficient of the upper level drops below that of the lower level, resulting in dropping the line source function relatively to the Planck function and strengthened lines, such that  $\Delta_{\rm NLTE} = -0.08$ and $-0.09$~dex, respectively.

In Paper~I, we applied a line-by-line differential NLTE and LTE approach, in the
sense that stellar line abundances were compared with individual abundances of
their solar counterparts. Here, we find that, similarly to the non-differential analysis, the differential abundances from the two ionisation stages: [Fe/H]$_{\rm I}$ = $-2.58\pm 0.11$ (YBK18) and $-2.62\pm 0.11$ (B18) and [Fe/H]$_{\rm II}$ = $-2.55\pm0.08$ agree within the error bars. 

\begin{figure*}
% \begin{center}
%\hspace{-0.1\linewidth}
% \centering}
% \hspace{0.2\linewidth}
  \resizebox{90mm}{!}{\includegraphics{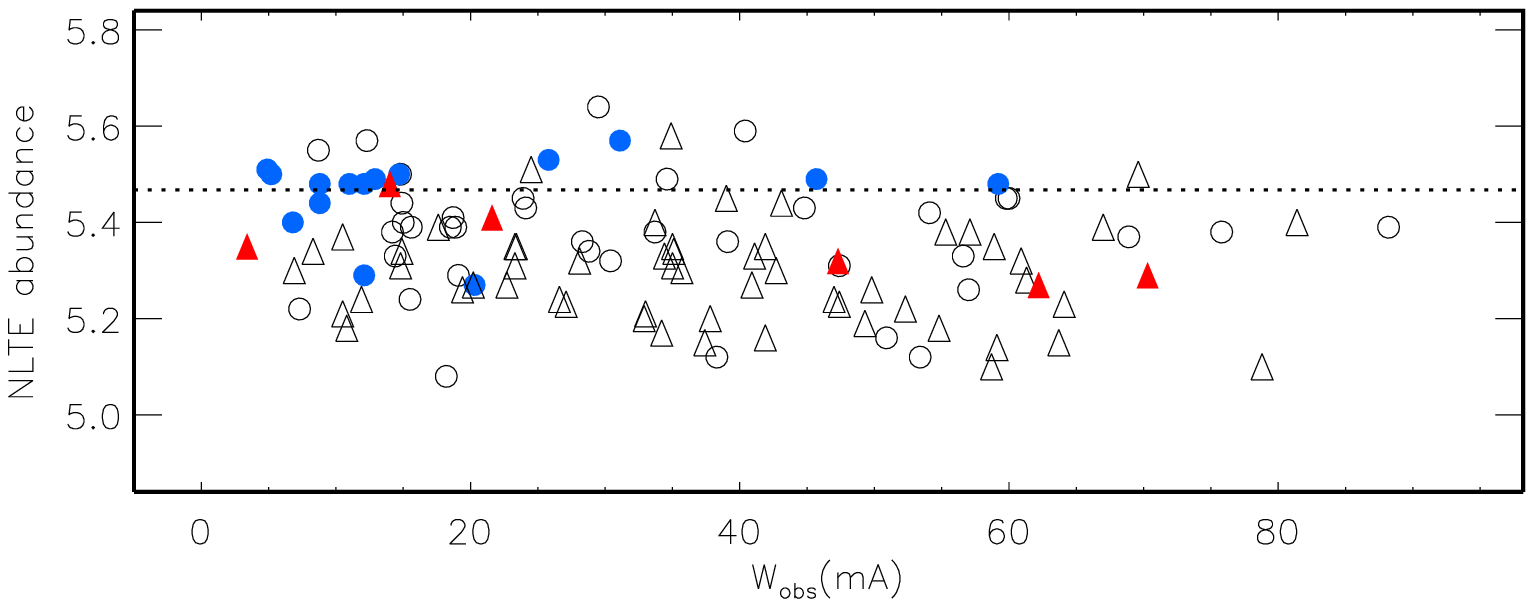}}
  \resizebox{90mm}{!}{\includegraphics{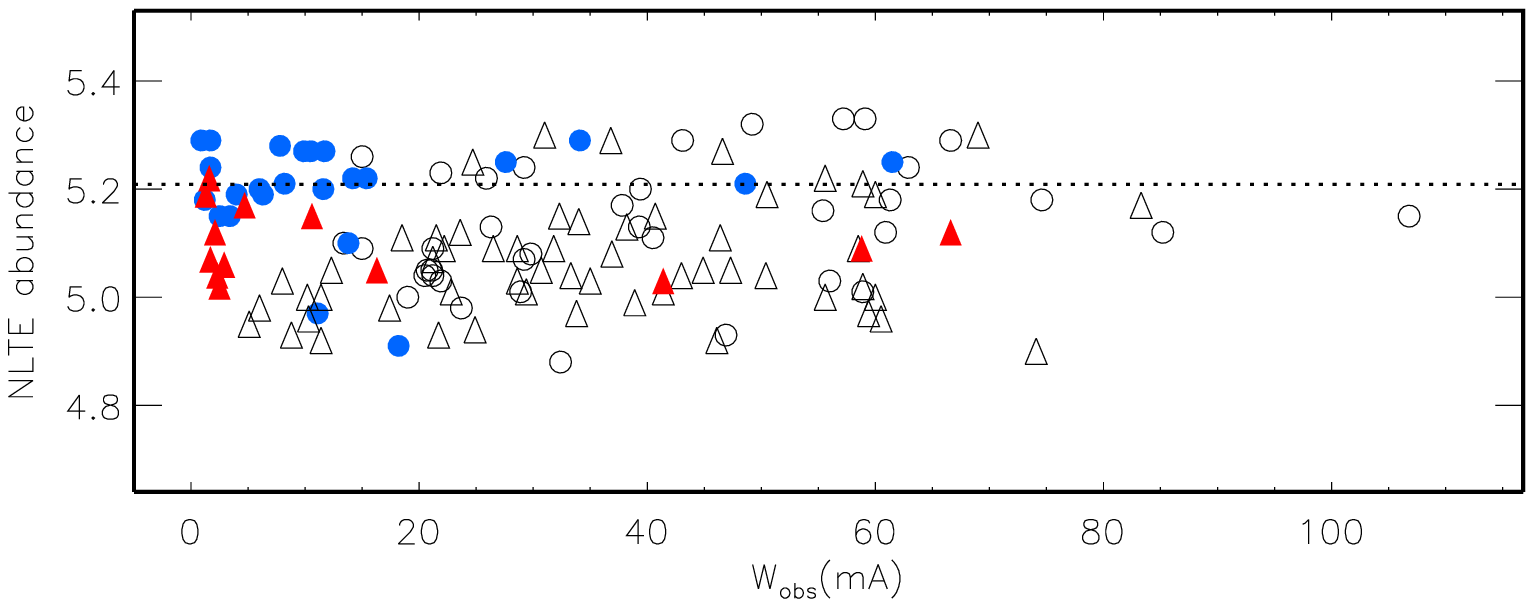}}
 \caption{NLTE abundances of HD~84937 (left panel) and HD~140283 (right panel) from lines of \ion{Fe}{i} (circles) and \ion{Fe}{ii} (triangles) in the YBK18+YBK19 scenario as a function of 
%excitation energy of the lower level, \Eexc, (left column) and 
observed equivalent width W$_{\rm obs}$. The filled and open symbols correspond to the visible and UV lines, respectively.
% in the (blue circles)  and \ion{Fe}{ii} in the visible (red triangles) and UV (open triangles) spectral range In the NLTE calculations, the \ion{Fe}{i} + \ion{H}{i} collisions are treated following B18 and the \ion{Fe}{ii} + \ion{H}{i} collisions following YBK19. 
In each panel, the dotted line indicates the mean abundance derived from the \ion{Fe}{i} visible lines.} 
 \label{fig:hd84937}
% \end{center}
 \end{figure*}

 \underline{HD~84937 and HD~140283.} 
 For the visible spectral range, we use spectra from the ESO UVESPOP survey \citep{2003Msngr.114...10B}. The line list is
 %together with the equivalent widths measured in high-quality  are
 taken from Paper~II, and it includes 12 lines of \ion{Fe}{i} and 7 lines of \ion{Fe}{ii} in HD~84937 and 20 and 14 lines in HD~140283. 
For lines of \ion{Fe}{i}, we adopt the same $gf$-values, as in Paper~II. The exceptions are \ion{Fe}{i} 5242, 5379, and 5662\,\AA, for which the data are taken from \citet{2017ApJ...848..125B} and \citet{2014ApJS..215...23D}. 
As for HD~122563, we use $gf$-values of RU98 for lines of \ion{Fe}{ii}, with the exceptions for \ion{Fe}{ii} 4923 and 5018\,\AA.
%for \ion{Fe}{ii} lines in the visible range. 
%It is worth noting that $gf$-values of \citet{RU} are used in analysis of  too.
  
  This study extends analysis to the UV spectral range by using
 high-quality HST/STIS spectra in the 1875 -- 3158~\AA\ range, with a quality factor (QF) per resel of 52 and 90 for HD~84937 and HD~140283, respectively. Observations are provided by Thomas Ayres at http://casa.colorado.edu/$\sim$ayres/ASTRAL/ within the ASTRAL project. 
 
From a vast number of the \ion{Fe}{i} lines in the UV spectrum of HD~140283, we selected those, which are not blended and have $gf$-values in \citet{1991JOSAB...8.1185O}. The average LTE abundance from these lines is referred below as $\eps{UV1}$. Then we appended some unblended lines with $gf$-values from \citet{Kurucz2009} and \citet{1988JPCRD..17S....F}, which provide the abundance consistent with $\eps{UV1}$, within 0.15~dex. For \ion{Fe}{i} 2487 and 2730\,\AA, laboratory $gf$-values are taken from \citet{2017ApJ...848..125B}. We find that the statistical error of the LTE abundance from \ion{Fe}{i} lines reduces from 0.36 to 0.14~dex, when moving from a  total list of 91 lines to the selected 42 lines. For 70 unblended lines of \ion{Fe}{ii}, with $gf$-values mostly from RU98, we obtain $\sigma \simeq$ 0.15~dex. For \ion{Fe}{ii} 2262 and 2268~\AA, $gf$-values are taken from \citet{KK} and for \ion{Fe}{ii} 2254~\AA\ from \citet{PGHcor}.
No further pre-selection was made. We use a common list of the UV lines for HD~84937 and HD~140283.  It includes 42 lines of \ion{Fe}{i} and 70 lines of \ion{Fe}{ii} in the 1968 -- 2990~\AA\ wavelength range (Tables~\ref{tab:lines} and A.1).

\begin{table*} %[htbp]
\caption{\label{tab:lines} LTE and NLTE abundances, $\eps{}$, from individual lines of \ion{Fe}{i} and \ion{Fe}{ii} in HD~84937, HD~140283, and 
HD~122563. 
The two sets of the NLTE abundances correspond to different collisional recipes, namely, B18 and YBK18. 
Observed equivalent widths, $W_{\rm obs}$, are given in m\AA. This table is available in its entirety in a machine-readable
form in the online version. A portion is shown here for guidance
regarding its form and content. } 
 \centering
\begin{tabular}{cccrccccrccccrccc}   %{ccl|cc|cc|cc|r}
\hline \noalign{\smallskip}
 $\lambda$,  & \Eexc, & $\log gf$ & \multicolumn{4}{c}{HD~84937} & & \multicolumn{4}{c}{HD~140283} & & \multicolumn{4}{c}{HD~122563} \\
\cline{4-7} \cline{9-12} \cline{14-17}
 \AA  &   eV     &      & $W_{\rm obs}$  & LTE  & B18 & YBK18 & & $W_{\rm obs}$  & LTE  & B18 & YBK18 & & $W_{\rm obs}$  & LTE  & B18 & YBK18 \\
\noalign{\smallskip} \hline \noalign{\smallskip}
\multicolumn{17}{c}{\ion{Fe}{i} lines} \\
% 4427.31 &  0.05 & -2.92 &       &       &       &       & &       &       &       &       & & 120.6 &  4.81 &  5.08 &  4.95 \\
 4445.48 &  0.09 & -5.44 &       &       &       &       & &       &       &       &       & &  14.6 &  4.82 &  4.97 &  4.94 \\
% 4574.72 &  2.28 & -2.97 &       &       &       &       & &       &       &       &       & &   6.8 &  4.78 &  4.94 &  5.00 \\
 4920.50 &  2.83 &  0.07 &  59.2 &  5.18 &  5.50 &  5.48 & &  61.5 &  4.91 &  5.23 &  5.25 & & 112.3 &  4.59 &  4.76 &  4.89 \\
 4994.13 &  0.92 & -2.96 &   8.8 &  5.28 &  5.45 &  5.44 & &  13.8 &  4.94 &  5.11 &  5.10 & &  77.6 &  4.80 &  4.91 &  4.88 \\
\hline \noalign{\smallskip}
\end{tabular}
\end{table*}

The NLTE calculations were performed for the YBK18+YBK19 and B18+YBK19 line formation scenarios. Figure~\ref{fig:hd84937} shows abundances from individual lines for YBK18+YBK19.
%with the \ion{Fe}{i} + \ion{H}{i} and \ion{Fe}{ii} + \ion{H}{i} collisions taken into account applying the data of B18 and YBK19, respectively. 

\begin{figure}
 \begin{center}
  \resizebox{90mm}{!}{\includegraphics{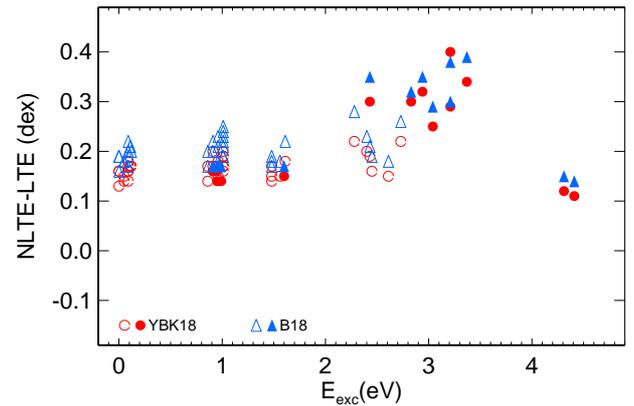}}
  \caption{NLTE abundance corrections for lines of \ion{Fe}{i} in the 6350/4.09/$-2.15$ model from calculations that include collisions with \ion{H}{i} according to YBK18 (circles) and B18 (triangles). The filled and open symbols correspond to the visible and UV lines, respectively. }
 \label{fig:dnlte84937}
 \end{center}
 \end{figure}

For both stars, the NLTE effects for lines of \ion{Fe}{ii} are minor, with $\Delta_{\rm NLTE} <$ 0.01~dex. In case of \ion{Fe}{i}, $\Delta_{\rm NLTE}$ is at the level of 0.15-0.2~dex for the \Eexc\ $<$ 2~eV lines, reaches a maximal value of $\sim 0.4$~dex for the lines arising from the \eu{z}{5}{D}{\circ}{} (\Eexc\ $\simeq$ 3.2~eV) and \eu{z}{5}{F}{\circ}{} (\Eexc\ $\simeq$ 3.4~eV)  levels and drops at \Eexc\ $>$ 4~eV  (Fig.~\ref{fig:dnlte84937} only for HD~84937). The difference in $\Delta_{\rm NLTE}$ between using the YBK18 and B18 data does not exceed 0.05~dex and 0.07~dex for every \ion{Fe}{i} line in HD~84937 and HD~140283, respectively. 

For each star, both in LTE and NLTE and for both \ion{Fe}{i} and \ion{Fe}{ii}, we obtain consistent within 0.08~dex abundances from the visible and UV lines. 
%The line-to-line scatter is larger for the UV than the visible lines. 
LTE leads to lower abundance from \ion{Fe}{i} compared with that from \ion{Fe}{ii}, by 0.08 to 0.16~dex for different spectral ranges and different stars. In NLTE, 
%increases and seems to overestimate abundances from \ion{Fe}{i}, such that 
the abundance difference $\eps{\rm FeI}- \eps{\rm FeII}$ is positive and ranges between 0.06 and 0.13~dex in different cases. 

\underline{Uncertainties in the derived iron abundances.}
For a given star, uncertainties in atmospheric parameters, applying different sources of the line atomic data, and different treatment of the NLTE line formation produce systematic shifts in the abundances derived from individual lines of a given chemical species.
We evaluate their effects on the \ion{Fe}{i}/\ion{Fe}{ii} ionisation equilibrium of the investigated stars, and the obtained results are summarised in Table~\ref{tab_uncert}. For comparison, the last string of Table~\ref{tab_uncert} displays the statistical errors due to line-to-line scatter, $\sigma_{\rm FeI - FeII} = \sqrt{\sigma_{\rm FeI}^2 + \sigma_{\rm FeII}^2}$. In test calculations, we vary $\Teff$, $\logg$, and $\xi_t$, by 50~K, 0.03~dex, and 0.2~\kms, respectively, taking into account quoted above errors of their measurements. 
%The uncertainty in microturbulent velocity is adopted to be .
%Effective temperatures and surface gravities were measured with an accuracy of 40~K to 50~K and 0.02 to 0.03~dex from
The string $gf$(MB09 -- RU98) of Table~\ref{tab_uncert} shows the changes in $\eps{\rm FeI} - \eps{\rm FeII}$, when replacing $gf$-values of RU98 with that of MB09 for the visible lines of  \ion{Fe}{ii}. 

Replacing photoionisation cross sections of \citet[][BLB2017]{2017A&A...606A.127B} with the older ones of \citet[][B1997]{1997A&AS..122..167B} leads to slightly larger NLTE effects for \ion{Fe}{i} lines in the 4600/1.4/$-2.5$ model atmosphere and nearly does not affect the NLTE results for the 6350/4.09/$-2.15$ and 5780/3.7/$-2.4$ models. 

In the model atmosphere of HD~122563, a use of the B18 rate coefficients for \ion{Fe}{i} + \ion{H}{i} collisions leads to smaller NLTE effects for \ion{Fe}{i} lines than that for the YBK18 data. In contrast, $\Delta_{\rm NLTE}$(B18) $> \Delta_{\rm NLTE}$(YBK18) for lines of \ion{Fe}{i} in the atmospheres of HD~84937 and HD~140283. However, the abundance shifts do not exceed 0.03~dex. 
 
For HD~122563, ignoring the \citet{1991JPhB...24L.127K} collision rates 
produces minor shifts in the NLTE abundances derived from \ion{Fe}{i} lines, of +0.02 and +0.03~dex, on average, in the B18 and YBK18 cases, respectively. Similarly small effects are found for HD~84937 and HD~140283. However, it is important to note that the abundance shifts have a different sign for the visible and the UV lines of \ion{Fe}{i}. For example, for HD~140283, ignoring the \citet {1991JPhB...24L.127K} collisions leads to the higher abundances from the visible lines, by 0.03~dex (YBK18) and $< 0.01$~dex (B18), while to the lower abundances from the UV lines, by 0.05 (YBK18) and 0.01~dex (B18). 
The abundance difference between the visible and the UV lines of \ion{Fe}{i} reduces, when the \citet{1991JPhB...24L.127K} collisions are taken into account. We note that similarly small effect of including the \citet{1991JPhB...24L.127K} collisions on the abundance determinations from lines of \ion{Mn}{i} is reported by \citet{2019arXiv190505200B}.

\begin{table*} %[htbp]
\caption{Shifts in $\eps{\rm FeI} - \eps{\rm FeII}$ caused by the uncertainties in atmospheric parameters, line atomic data, and NLTE treatment. \label{tab_uncert}}
 \centering
 \begin{tabular}{lccc}
\hline\hline \noalign{\smallskip}
     & HD~122563  & HD~84937   & HD~140283 \\
\noalign{\smallskip} \hline \noalign{\smallskip}
$\Delta \Teff$ = $-50$~K & $-0.11$  & $-0.04$  &  $-0.03$ \\
$\Delta \logg$ = +0.03   & $-0.04$  & $-0.01$  & $-0.01$ \\
$\Delta \xi_t$ = $-0.2$~\kms & +0.04 & $-0.02$  &  $-0.01$ \\
$gf$(MB09 -- RU98)  &  +0.07  & +0.04  &  +0.08  \\  
ph-ion (BLB2017 -- B1997) & $-0.02$ & $ < 0.01$ & $ < 0.01$ \\
\ion{Fe}{i} + \ion{H}{i} (B18 -- YBK18) & $-0.02$ & +0.03 & +0.03 \\
no Kaulakys' collisions, YBK18, vis & +0.03  & +0.02 & +0.03 \\
no Kaulakys' collisions, YBK18, UV  & --           & $ -0.01$ & $ -0.05$ \\
no Kaulakys' collisions, B18, vis+UV & +0.02    & +0.01 & $ < 0.01$ \\
line-to-line scatter, YBK18+YBK19, $\sigma_{\rm FeI - FeII}$ & ~~0.14 & ~~0.16 & ~~0.15 \\
\noalign{\smallskip}\hline \noalign{\smallskip}
\end{tabular}
\end{table*}

It can be seen that, for cool giants, 
%the uncertainty in $\Teff$ remains the major source of the uncertainties in the derived abundances and
achieving the \ion{Fe}{i}/\ion{Fe}{ii} ionisation equilibrium depends, in the first turn, on an accuracy of $\Teff$. Next important source of the abundance uncertainties is $gf$-values. For example, applying $gf$-values of MB09 could remove the NLTE abundance difference $\eps{\rm FeI}- \eps{\rm FeII}$ completely for HD~122563. We therefore call once more on atomic spectroscopists for further laboratory and theoretical works to improve $gf$-values of the \ion{Fe}{i} and, in particular, \ion{Fe}{ii} lines used in abundance analysis. Taking into account the obtained statistical abundance errors, we conclude that none of the systematic effects can destroy the \ion{Fe}{i}/\ion{Fe}{ii} ionisation equilibrium for the investigated halo benchmark stars.

\subsection{VMP giants in the dwarf spheroidal galaxies}\label{sec:dSphs}

%Another useful tool for testing recent data on the \ion{Fe}{i} + \ion{H}{i} and \ion{Fe}{ii} + \ion{H}{i} collisions is provided by the VMP stars in the dwarf spheroidal galaxies (dSphs) with well determined distances.
Our sample of the dSph stars covers the $-4 \le$ [Fe/H] $\le -1.5$ metallicity range that is useful for testing a updated model atom of \ion{Fe}{i-ii}, because the NLTE effects for \ion{Fe}{i} lines depend on stellar metallicity. We use the same line list as in \citet{dsph_parameters} and \citet{Pakhomov_boo1}. The same atomic parameters are adopted for lines of \ion{Fe}{i}, while, by analogy with analysis of the halo benchmark stars in Sect.~\ref{sec:halo}, $gf$-values of RU98 are employed for the \ion{Fe}{ii} lines.
%from , which From the cited papers we take the atmospheric parameters, the line list and , and the equivalent widths, which were measured in the high-resolution stellar spectra. The only difference is that we use here 
%with our previous studies of the dSph stars is for which  the data of, 
%in order to have a common base 

 \begin{figure}
 \begin{center}
  \resizebox{90mm}{!}{\includegraphics{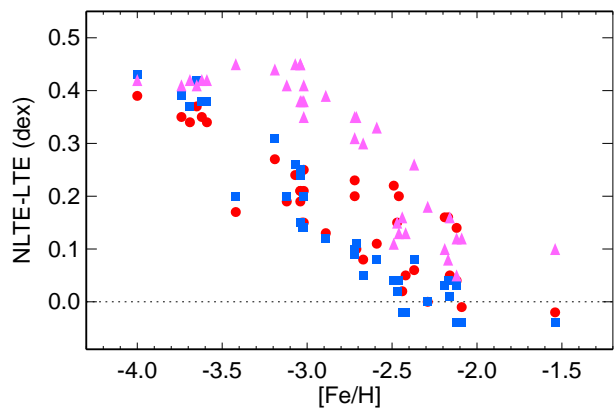}}
  \caption{NLTE abundance corrections for \ion{Fe}{i} 5216\,\AA\ (\Eexc\ = 1.61~eV, red  circles), 6191\,\AA\ (\Eexc\ = 2.43~eV, blue squares), and 5615\,\AA\ (\Eexc\ = 3.33~eV, magenda triangles)
%  and \ion{Fe}{ii} 5018\,\AA\ (\Eexc\ = 2.89~eV, open circles) and 5197\,\AA\ (\Eexc\ = 3.23~eV, open triangles)
in the dSph stars from calculations with the YBK18 data.}
 \label{fig:dsphs_dnlte}
 \end{center}
 \end{figure}

For each star, the NLTE calculations were performed with the YBK18+YBK19 collisional recipe. For lines of \ion{Fe}{ii}, the NLTE effects are minor at [Fe/H] $> -3.7$, such that $\Delta_{\rm NLTE} \le$ 0.02~dex, in absolute value. The exceptions are \ion{Fe}{ii} 4923 and 5018\,\AA, for which $\Delta_{\rm NLTE}$ is negative and can reach $-0.09$~dex. At the lowest metallicity of our sample, [Fe/H] $\sim -4$, the NLTE abundance corrections become positive for all the \ion{Fe}{ii} lines and reach +0.12~dex for 5197\,\AA\ and +0.06~dex for 5018\,\AA.

Figure~\ref{fig:dsphs_dnlte} displays the NLTE abundance corrections for three \ion{Fe}{i} lines with different \Eexc.
%\ and the two \ion{Fe}{ii} lines of different strength. 
The departures from LTE for \ion{Fe}{i} grow towards lower metallicity, for example, for \ion{Fe}{i} 5216\,\AA, from $\Delta_{\rm NLTE}$ = 0.13~dex in the 4275/0.65/$-2.08$ model to $\Delta_{\rm NLTE}$ = 0.39~dex in the 4800/1.56/$-4.0$ model. The NLTE corrections are larger, as a rule, for the higher than the lower excitation line. For example, in the 4850/2.05/$-2.96$ model, $\Delta_{\rm NLTE}$ = 0.52, 0.25, and 0.20~dex for \ion{Fe}{i} 5615\,\AA\ (\Eexc\ = 3.33~eV), 6191\,\AA\ (\Eexc\ = 2.43~eV), and 5216\,\AA\ (\Eexc\ = 1.61~eV), respectively. Although the discrepancy between different lines reduces towards lower metallicity.
Tables~\ref{tab:dsphs} and A.2 (online material) present the average LTE and NLTE abundances of the dSph stars. 

\begin{table*} %[htbp]
\caption{\label{tab:dsphs} Iron NLTE abundances, $\eps{}$, of the dSph stars from calculations with the YBK18+YBK19 recipe. This table is available in its entirety in a machine-readable
form in the online version. A portion is shown here for guidance
regarding its form and content.} 
 \centering
\begin{tabular}{lccccrrcccccc}   %{ccl|cc|cc|cc|r}
\hline \noalign{\smallskip}
ID & $\Teff$ & $\logg$ & [Fe/H] & $\xi_t$ & \multicolumn{2}{c}{N$_l$} & & \multicolumn{2}{c}{LTE} & & \multicolumn{2}{c}{NLTE} \\
\cline{6-7} \cline{9-10} \cline{12-13}
%\cline{6-10}
   &   K     & CGS     &        &  \kms   & \ion{Fe}{i} & \ion{Fe}{ii} & & \ion{Fe}{i} & \ion{Fe}{ii} & & \ion{Fe}{i} & \ion{Fe}{ii}  \\
\noalign{\smallskip} \hline \noalign{\smallskip}
\multicolumn{1}{c}{(1)} & (2) & (3) & (4) & (5) & (6) & (7) & & (8) & (9) & & (10) & (11)\\ 
 \noalign{\smallskip}  \hline \noalign{\smallskip} 
Scl ET0381     & 4570 & 1.17 & -2.19 & 1.7 & 74 &  9 & &  5.14(0.17)$^1$ &  5.42(0.09) & &  5.29(0.16) &  5.42(0.09) \\
Scl002\_06     & 4390 & 0.68 & -3.11 & 2.3 & 69 &  4 & &  4.12(0.17) &  4.50(0.10) & &  4.42(0.14) &  4.48(0.09) \\
Scl03\_059     & 4530 & 1.08 & -2.88 & 1.9 & 91 &  4 & &  4.43(0.17) &  4.73(0.10) & &  4.72(0.14) &  4.69(0.08) \\
\hline \noalign{\smallskip}
\multicolumn{13}{l}{$^1$ The numbers in  parentheses are the dispersions in the single line measurements around the mean.} \\
\end{tabular}
\end{table*}

 \begin{figure}
 \begin{center}
  \resizebox{90mm}{!}{\includegraphics{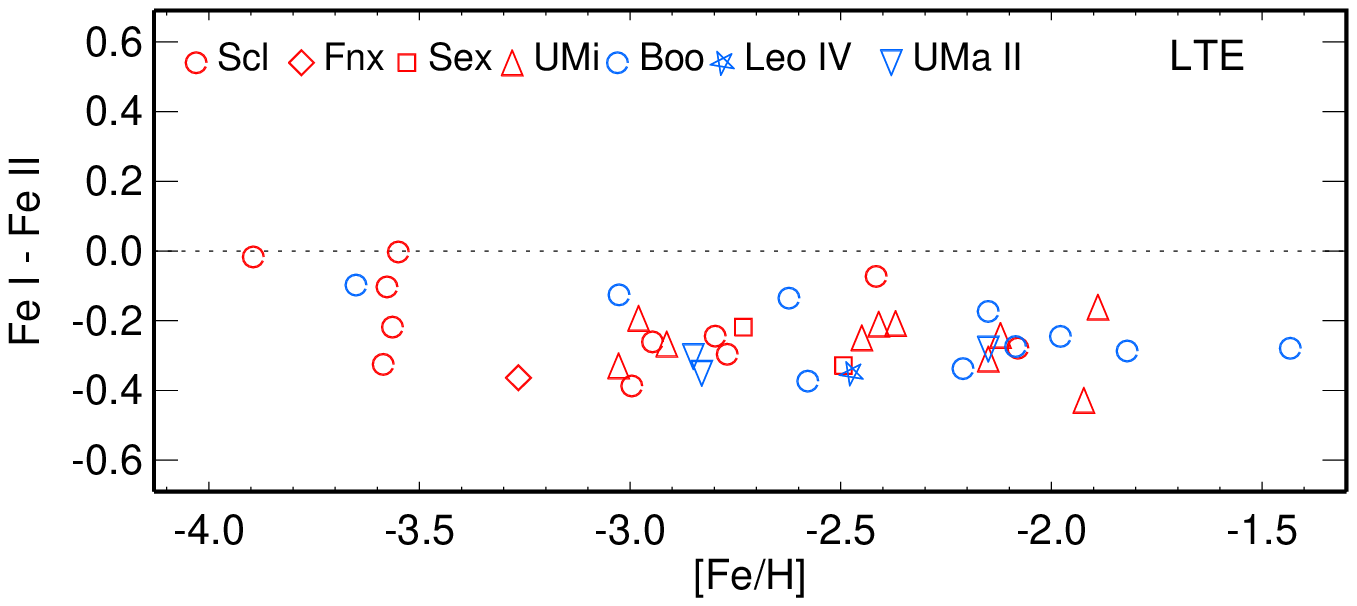}}
  \resizebox{90mm}{!}{\includegraphics{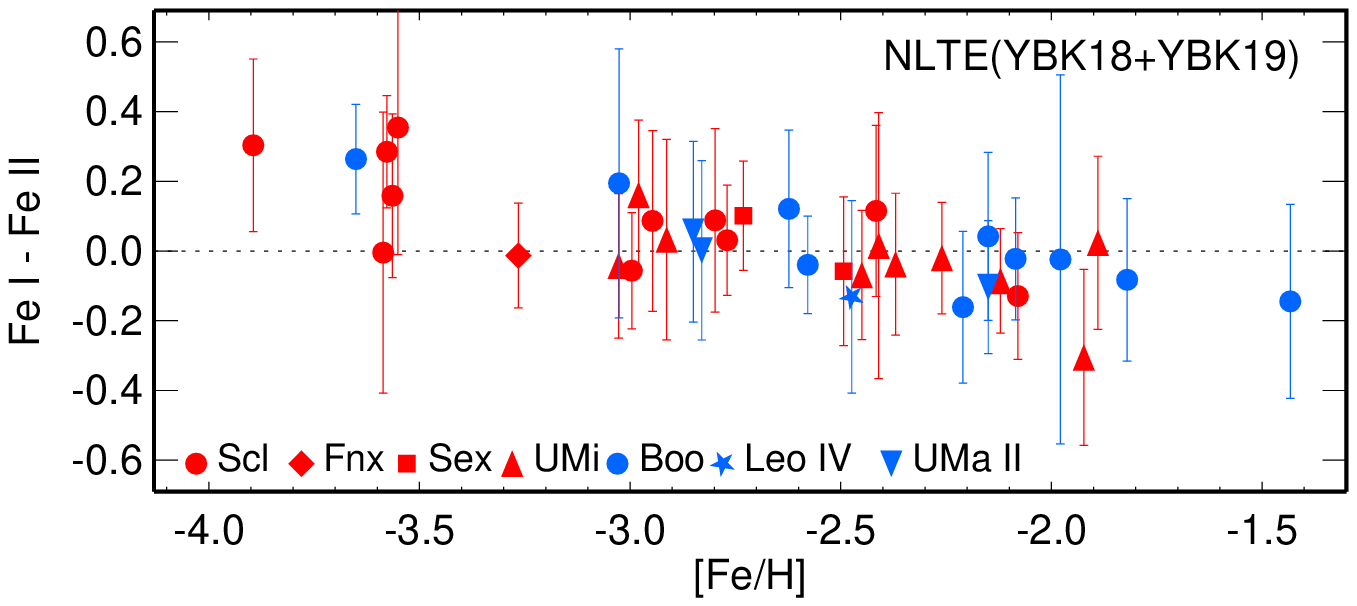}}
  \caption{LTE (top panel) and NLTE (YBK18+YBK19, bottom panel) abundance differences between \ion{Fe}{i} and \ion{Fe}{ii}, $\eps{FeI}$ -- $\eps{FeII}$, in the stars in Sculptor (red circles), Ursa Minor (red triangles), Fornax (red rhombi), Sextans (red squares), Bo{\"o}tes~I (blue circles), UMa~II (blue inverted triangles), and Leo~IV (blue 5 pointed star) dSphs. The error bars corresponds to $\sigma_{\rm FeI - FeII}$. }
 \label{fig:dsphs}
 \end{center}
 \end{figure}

The abundance differences between \ion{Fe}{i} and \ion{Fe}{ii} are displayed in Fig.~\ref{fig:dsphs}. %As was pointed out by  
Under the LTE assumption, $\eps{FeI}$ is systematically lower than $\eps{FeII}$, by $-0.27\pm$0.08~dex, on average. In the NLTE calculations, an abundance discrepancy is largely removed in the [Fe/H] $> -3.5$ regime, with the mean $\eps{FeI}$ -- $\eps{FeII}$ = $-0.01\pm$0.10. At the lower metallicity, in five of six stars, abundances from \ion{Fe}{i} lines are higher that those from \ion{Fe}{ii}, by up to 0.35~dex. As discussed by \citet{dsph_parameters}, this can be due to overestimated effective temperatures that were determined for these stars from photometric colours. From a carefull analysis of a sample of the UMP stars, \citet{2019MNRAS.485.3527S} recommend to use as many as possible photometric and spectroscopic indicators of $\Teff$ and $\logg$, in order to improve an accuracy of derived atmospheric parameters. In this study, we did not aim to revise $\Teff$'s of our sample stars.

%or using only two lines of \ion{Fe}{ii} (at 4923 and 5018~\AA).

Analysis of the \ion{Fe}{i} /\ion{Fe}{ii} ionisation equilibrium in the same dSph stars by \citet{dsph_parameters} was different from the present one in the two aspects: collisions with \ion{H}{i} were treated with the Drawinian rates and abundances from the \ion{Fe}{ii} lines were derived with $gf$-values of \citet{RU}, which were corrected by $+0.11$~dex, following the recommendation of \citet{Grevesse1999}. In order to achieve consistent abundances from the two ionisation stages for the stars at [Fe/H] $> -3.5$, \citet{dsph_parameters} applied a scaling factor of \kH\ = 0.5 to the Drawinian rates. With the quantum mechanical rate coefficients from  YBK18, we calculate larger NLTE abundance corrections for \ion{Fe}{i} lines, by  0.1~dex, on average. Therefore, a use of the scaled (\kH\ = 0.5) Drawinian rates together with the corrected (by $+0.11$~dex) $gf$-values of \citet{RU} provides, on average, the same abundance difference between \ion{Fe}{i} and \ion{Fe}{ii}, as that for
accurate collisional data together with original $gf$-values of \citet{RU}. 

\section{Comparison with other studies}\label{Sect:comparison}

For the halo benchmark stars discussed in Sect.~\ref{sec:halo}, the iron abundance analyses were performed by \citet[][hereafter, ALA16]{2016MNRAS.463.1518A} based on detailed 3D NLTE radiative transfer calculations using 3D hydrodynamic {\sc STAGGER} model atmospheres. They  treated \ion{Fe}{i} + \ion{H}{i} collisions with quantum-mechanical rate coefficients from \citet{2016PhRvA..93d2705B}. We do not compare here the results for HD~140283 because of large discrepancy in $\Teff$ between this study (5780~K) and ALA16 (5591~K). 
%The latter value is not supported by the most recent measurements of \citet{2018MNRAS.475L..81K}, who determined $\Teff$ = 5787$\pm$48~K.
For HD~84937, ALA16 adopted $\Teff$ = 6356~K, $\logg$ = 4.06, and [Fe/H] = $-2.0$, 
which are close to ours. Our LTE abundances from lines of both \ion{Fe}{i} and \ion{Fe}{ii} in the visible spectral range agree with the corresponding 1D LTE abundances of ALA16, within 0.04~dex. We obtain slightly stronger NLTE effects for \ion{Fe}{i} lines, with the mean NLTE -- LTE abundance difference of +0.22(YBK18) / 0.24~dex (B18), while the 1D NLTE analysis of ALA16 results in +0.17~dex. 

In order to compare the results for HD~122563, we take the same atmospheric parameters, $\Teff$ = 4600~K, $\logg$ = 1.6, and [Fe/H] = $-2.5$, as that in ALA16. In LTE, we obtain higher abundances than those of ALA16, by 0.13 and 0.10~dex for \ion{Fe}{i} and \ion{Fe}{ii}, respectively. The NLTE -- LTE abundance differences  amount to 0.15 (YBK18) and 0.13~dex (B18) for \ion{Fe}{i}, while ALA16 reported 0.09~dex. The discrepancies in both LTE and NLTE abundances can be due to using different line lists.

Interestingly, the 3D NLTE analyses of ALA16 lead to the higher abundances compared with the 1D NLTE ones, by a very similar amount for \ion{Fe}{i} and \ion{Fe}{ii} in a given star: by 0.12 and 0.10~dex, respectively, in HD~84937 and by 0.08 and 0.07~dex in HD~122563. This leads us to think that the NLTE abundance discrepancy of 0.1~dex between \ion{Fe}{i} and \ion{Fe}{ii} obtained in this study for the visible lines in HD~84937 and HD~140283 could not be removed, when applying the 3D NLTE analysis.

For HD~84937, the \ion{Fe}{i} /\ion{Fe}{ii} ionisation equilibrium based on not only visible, but also UV spectra was checked earlier by \citet[][hereafter, S16]{2016ApJ...817...53S} and \citet[][hereafter, R18]{2018ApJ...860..125R} under the LTE assumption. Both studies used the VLT/UVES spectrum in the visible range, but different  STIS UV spectra, with a resolution of R = 25\,000 and 114\,000, respectively. 
Despite employing rather different atmospheric parameters, namely, $\Teff$/$\logg$ = 6300~K/4.00 (S16) and 6418~K/4.16 (R18), these two papers reported close together abundances, and 
each of them obtained a perfect agreement between the two ionisation stages, \ion{Fe}{i} and  \ion{Fe}{ii}. 
From 446 lines of \ion{Fe}{i} and 105 lines of \ion{Fe}{ii}, S16 derived $\eps{\rm FeI}$ = 5.20 and $\eps{\rm FeII}$ = 5.19. Using the solar abundances presented by S16 in their Table~1, we calculate identical values: [Fe/H]$_{\rm I}$ = [Fe/H]$_{\rm II} = -2.32$. In their Table~8, R18 present [Fe/H]$_{\rm I} = -2.26$ from 164 lines with \Eexc\ $> 1.2$~eV and [Fe/H]$_{\rm II} = -2.23$ from 27 lines.

Our list of the UV lines has 20 lines of \ion{Fe}{i} and 6 lines of \ion{Fe}{ii} in common with S16, while no common line with R18. By averaging our results from the visible and UV lines, we obtain the LTE abundance $\eps{\rm FeI}$ = 5.23, in line with that of S16 and R18. Our $\eps{\rm FeII}$ is higher than that of S16, by 0.11~dex, and agrees within 0.02~dex with that of R18. A discrepancy with S16 can be due to different line lists and different $gf$-values. When using the same $gf$-values, as in S16 for common lines, we obtain a difference of less than 0.01~dex in $\eps{\rm FeII}$ between this study and S16.

\section{Conclusions}\label{Sect:Conclusions}

In this study, the \ion{Fe}{i} -\ion{Fe}{ii} model atom of \citet{mash_fe} is updated by implementing photoionisation cross sections for the \ion{Fe}{i} levels from  \citet{2017A&A...606A.127B}, electron-impact excitation data of \citet{2017A&A...606A.127B} for the \ion{Fe}{i} transitions, and by including inelastic \ion{Fe}{i} + \ion{H}{i} and \ion{Fe}{ii} + \ion{H}{i} collisions with rate coefficients from quantum-mechanical calculations of \citet[][\ion{Fe}{i}]{2018A&A...612A..90B}, \citet[][\ion{Fe}{i}]{2018CP....515..369Y}, and \citet[][\ion{Fe}{ii}]{2019MNRAS.483.5105Y}. 
We also implement the \citet{1991JPhB...24L.127K}' collisions applying rate coefficients from calculations of \citet{2017ascl.soft01005B}.

Using classical 1D model atmospheres, we inspect the effect of \ion{H}{i} collisions and their different treatment on the derived abundances and the \ion{Fe}{i} /\ion{Fe}{ii} ionisation equilibrium of the three Galactic halo benchmark stars and a sample of 38 VMP giants in the dwarf galaxies with well known distances. For each star, the average NLTE abundances from \ion{Fe}{i} lines obtained using the data of \citet{2018CP....515..369Y} and \citet{2018A&A...612A..90B} differ by no more than 0.03~dex. Although there exist discrepancies in the NLTE abundance corrections for individual lines. In most cases, they do not exceed  0.1~dex. The exceptions are the \ion{Fe}{i} 4427, 4920, and 5324\,\AA\ lines in HD~122563, for which the difference in $\Delta_{\rm NLTE}$ amounts to 0.17, 0.13, and 0.15~dex, respectively.
Further theoretical works are needed to improve the theory of inelastic \ion{Fe}{i} + \ion{H}{i} collisions. 

The obtained results can be summarized as follows.
\begin{itemize}
\item Collisions with \ion{H}{i} serve as efficient thermalisation source for \ion{Fe}{ii}, such that the NLTE abundance corrections for \ion{Fe}{ii} lines do not exceed 0.02~dex, in absolute value, at [Fe/H] $\gtrsim -3$ and can reach +0.06~dex at [Fe/H] $\sim -4$. Thus, in a broad metallicity range, except the ultra-metal poor stars, lines of \ion{Fe}{ii} can be safely used under the LTE assumption as an abundance indicator.
\item For the VMP giants, the NLTE abundances from the two ionisation stages agree within the error bars, with $\eps{FeI}$ -- $\eps{FeII}$ = $-0.07$ (YBK18) and $-0.09$ (B18) for HD~122563 and $-0.01\pm$0.10 for 32 stars in the dSphs at [Fe/H] $> -3.5$. For comparison, the corresponding LTE abundance differences amount to $-0.25$~dex and $-0.27\pm$0.08~dex.
\item Using the 1960~\AA\ to 6460~\AA\ spectra, we obtain an abundance discrepancy of 0.10 (YBK18) / 0.13 (B18) and 0.09 (YBK18) / 0.12 (B18) between \ion{Fe}{i} and \ion{Fe}{ii} in a VMP turn-off star HD~84937 and a VMP subgiant HD~140283, respectively. For either \ion{Fe}{i} or \ion{Fe}{ii} in each star, abundances from the visible and UV lines are found to be consistent within 0.08~dex. 
\end{itemize}

\begin{acknowledgements}
We are grateful to Paul Barklem and Manuel Bautista for providing the rate coefficients for \ion{Fe}{i} + \ion{H}{i} collisions and photoionisation cross sections for \ion{Fe}{i} publicly available.
The authors thank the Russian Science Foundation (grant 17-13-01144) for a partial support of this study. We made use of the ESO UVESPOP and the ASTRAL spectral archives and the ADS\footnote{http://adsabs.harvard.edu/abstract\_service.html}, SIMBAD\footnote{http://simbad.u-strasbg.fr/simbad/}, and VALD databases.
\end{acknowledgements}

\bibliography{atomic_data,mashonkina,nlte,references,fe2}

\begin{thebibliography}{66}
\expandafter\ifx\csname natexlab\endcsname\relax\def\natexlab#1{#1}\fi

\bibitem[{{Amarsi} {et~al.}(2016){Amarsi}, {Lind}, {Asplund}, {Barklem}, \&
  {Collet}}]{2016MNRAS.463.1518A}
{Amarsi}, A.~M., {Lind}, K., {Asplund}, M., {Barklem}, P.~S., \& {Collet}, R.
  2016, \mnras, 463, 1518

\bibitem[{{Bagnulo} {et~al.}(2003){Bagnulo}, {Jehin}, {Ledoux}, {Cabanac},
  {Melo}, {Gilmozzi}, \& {ESO Paranal Science Operations
  Team}}]{2003Msngr.114...10B}
{Bagnulo}, S., {Jehin}, E., {Ledoux}, C., {et~al.} 2003, The Messenger, 114, 10

\bibitem[{{Bailer-Jones} {et~al.}(2018){Bailer-Jones}, {Rybizki}, {Fouesneau},
  {Mantelet}, \& {Andrae}}]{2018AJ....156...58B}
{Bailer-Jones}, C.~A.~L., {Rybizki}, J., {Fouesneau}, M., {Mantelet}, G., \&
  {Andrae}, R. 2018, \aj, 156, 58

\bibitem[{{Barklem}(2016)}]{2016PhRvA..93d2705B}
{Barklem}, P.~S. 2016, \pra, 93, 042705

\bibitem[{{Barklem}(2017)}]{2017ascl.soft01005B}
{Barklem}, P.~S. 2017, {KAULAKYS: Inelastic collisions between hydrogen atoms
  and Rydberg atoms}

\bibitem[{{Barklem}(2018)}]{2018A&A...612A..90B}
{Barklem}, P.~S. 2018, \aap, 612, A90

\bibitem[{{Bautista}(1997)}]{1997A&AS..122..167B}
{Bautista}, M.~A. 1997, \aaps, 122, 167

\bibitem[{{Bautista} {et~al.}(2017){Bautista}, {Lind}, \&
  {Bergemann}}]{2017A&A...606A.127B}
{Bautista}, M.~A., {Lind}, K., \& {Bergemann}, M. 2017, \aap, 606, A127

\bibitem[{{Bautista} \& {Pradhan}(1996)}]{1996A&AS..115..551B}
{Bautista}, M.~A. \& {Pradhan}, A.~K. 1996, \aaps, 115, 551

\bibitem[{{Bautista} \& {Pradhan}(1998)}]{1998ApJ...492..650B}
{Bautista}, M.~A. \& {Pradhan}, A.~K. 1998, \apj, 492, 650

\bibitem[{{Belmonte} {et~al.}(2017){Belmonte}, {Pickering}, {Ruffoni}, {Den
  Hartog}, {Lawler}, {Guzman}, \& {Heiter}}]{2017ApJ...848..125B}
{Belmonte}, M.~T., {Pickering}, J.~C., {Ruffoni}, M.~P., {et~al.} 2017, \apj,
  848, 125

\bibitem[{{Bensby} {et~al.}(2014){Bensby}, {Feltzing}, \& {Oey}}]{Bensby2014}
{Bensby}, T., {Feltzing}, S., \& {Oey}, M.~S. 2014, \aap, 562, A71

\bibitem[{{Bergemann} {et~al.}(2019){Bergemann}, {Gallagher}, {Eitner},
  {Bautista}, {Collet}, {Yakovleva}, {Mayriedl}, {Plez}, {Carlsson},
  {Leenaarts}, {Belyaev}, \& {Hansen}}]{2019arXiv190505200B}
{Bergemann}, M., {Gallagher}, A.~J., {Eitner}, P., {et~al.} 2019, arXiv
  e-prints, arXiv:1905.05200

\bibitem[{{Bergemann} {et~al.}(2012){Bergemann}, {Lind}, {Collet}, {Magic}, \&
  {Asplund}}]{Bergemann_fe_nlte}
{Bergemann}, M., {Lind}, K., {Collet}, R., {Magic}, Z., \& {Asplund}, M. 2012,
  \mnras, 427, 27

\bibitem[{{Bergemann} {et~al.}(2014){Bergemann}, {Ruchti}, {Serenelli},
  {Feltzing}, {Alves-Brito}, {Asplund}, {Bensby}, {Gruyters}, {Heiter},
  {Hourihane}, {Korn}, {Lind}, {Marino}, {Jofre}, {Nordlander}, {Ryde},
  {Worley}, {Gilmore}, {Randich}, {Ferguson}, {Jeffries}, {Micela},
  {Negueruela}, {Prusti}, {Rix}, {Vallenari}, {Alfaro}, {Allende Prieto},
  {Bragaglia}, {Koposov}, {Lanzafame}, {Pancino}, {Recio-Blanco}, {Smiljanic},
  {Walton}, {Costado}, {Franciosini}, {Hill}, {Lardo}, {de Laverny}, {Magrini},
  {Maiorca}, {Masseron}, {Morbidelli}, {Sacco}, {Kordopatis}, \& {Tautvai{\v
  s}ien{\.e}}}]{Bergemann2014A&A...565A..89B}
{Bergemann}, M., {Ruchti}, G.~R., {Serenelli}, A., {et~al.} 2014, \aap, 565,
  A89

\bibitem[{{Boyarchuk} {et~al.}(1985){Boyarchuk}, {Lyubimkov}, \&
  {Sakhibullin}}]{1985Ap.....22..203B}
{Boyarchuk}, A.~A., {Lyubimkov}, L.~S., \& {Sakhibullin}, N.~A. 1985,
  Astrophysics, 22, 203

\bibitem[{{Bridges}(1973)}]{Bridges_fe2}
{Bridges}, J.~M. 1973, in Phenomena in Ionized Gases, Eleventh International
  Conference, ed. {I.~{\v S}toll}, 418--+, (B)

\bibitem[{{Butler} \& {Giddings}(1985)}]{detail}
{Butler}, K. \& {Giddings}, J. 1985, Newsletter on the analysis of astronomical
  spectra, No. 9, University of London

\bibitem[{{Creevey} {et~al.}(2019){Creevey}, {Grundahl}, {Th{\'e}venin},
  {Corsaro}, {Pall{\'e}}, {Salabert}, {Pichon}, {Collet}, {Bigot}, {Antoci}, \&
  {Andersen}}]{2019arXiv190202657C}
{Creevey}, O., {Grundahl}, F., {Th{\'e}venin}, F., {et~al.} 2019, arXiv
  e-prints

\bibitem[{{Creevey} {et~al.}(2012){Creevey}, {Th{\'e}venin}, {Boyajian},
  {Kervella}, {Chiavassa}, {Bigot}, {M{\'e}rand}, {Heiter}, {Morel}, {Pichon},
  {Mc Alister}, {ten Brummelaar}, {Collet}, {van Belle}, {Coud{\'e} du
  Foresto}, {Farrington}, {Goldfinger}, {Sturmann}, {Sturmann}, \&
  {Turner}}]{2012A&A...545A..17C}
{Creevey}, O.~L., {Th{\'e}venin}, F., {Boyajian}, T.~S., {et~al.} 2012, \aap,
  545, A17

\bibitem[{{Den Hartog} {et~al.}(2014){Den Hartog}, {Ruffoni}, {Lawler},
  {Pickering}, {Lind}, \& {Brewer}}]{2014ApJS..215...23D}
{Den Hartog}, E.~A., {Ruffoni}, M.~P., {Lawler}, J.~E., {et~al.} 2014, \apjs,
  215, 23

\bibitem[{{Drawin}(1968)}]{Drawin1968}
{Drawin}, H.-W. 1968, Zeitschrift fur Physik, 211, 404

\bibitem[{{Drawin}(1969)}]{Drawin1969}
{Drawin}, H.~W. 1969, Zeitschrift fur Physik, 225, 483

\bibitem[{{Fuhr} {et~al.}(1988){Fuhr}, {Martin}, \&
  {Wiese}}]{1988JPCRD..17S....F}
{Fuhr}, J.~R., {Martin}, G.~A., \& {Wiese}, W.~L. 1988, Journal of Physical and
  Chemical Reference Data, 17

\bibitem[{{Gaia Collaboration} {et~al.}(2018){Gaia Collaboration}, {Brown},
  {Vallenari}, {Prusti}, {de Bruijne}, {Babusiaux}, {Bailer-Jones}, {Biermann},
  {Evans}, {Eyer}, \& et~al.}]{2018A&A...616A...1G}
{Gaia Collaboration}, {Brown}, A.~G.~A., {Vallenari}, A., {et~al.} 2018, \aap,
  616, A1

\bibitem[{{Gehren} {et~al.}(2001){Gehren}, {Butler}, {Mashonkina}, {Reetz}, \&
  {Shi}}]{2001A&A...366..981G}
{Gehren}, T., {Butler}, K., {Mashonkina}, L., {Reetz}, J., \& {Shi}, J. 2001,
  \aap, 366, 981

\bibitem[{{Gigas}(1986)}]{1986A&A...165..170G}
{Gigas}, D. 1986, \aap, 165, 170

\bibitem[{{Gratton} {et~al.}(1999){Gratton}, {Carretta}, {Eriksson}, \&
  {Gustafsson}}]{Gratton1999}
{Gratton}, R.~G., {Carretta}, E., {Eriksson}, K., \& {Gustafsson}, B. 1999,
  \aap, 350, 955

\bibitem[{{Grevesse} \& {Sauval}(1999)}]{Grevesse1999}
{Grevesse}, N. \& {Sauval}, A.~J. 1999, \aap, 347, 348

\bibitem[{{Grupp} {et~al.}(2009){Grupp}, {Kurucz}, \& {Tan}}]{Grupp2009}
{Grupp}, F., {Kurucz}, R.~L., \& {Tan}, K. 2009, \aap, 503, 177

\bibitem[{Gustafsson {et~al.}(2008)Gustafsson, Edvardsson, Eriksson, Jorgensen,
  Nordlund, \& Plez}]{Gustafssonetal:2008}
Gustafsson, B., Edvardsson, B., Eriksson, K., {et~al.} 2008, A\&A, 486, 951

\bibitem[{{Karovicova} {et~al.}(2018){Karovicova}, {White}, {Nordlander},
  {Lind}, {Casagrande}, {Ireland}, {Huber}, {Creevey}, {Mourard}, {Schaefer},
  {Gilmore}, {Chiavassa}, {Wittkowski}, {Jofr{\'e}}, {Heiter}, {Th{\'e}venin},
  \& {Asplund}}]{2018MNRAS.475L..81K}
{Karovicova}, I., {White}, T.~R., {Nordlander}, T., {et~al.} 2018, \mnras, 475,
  L81

\bibitem[{{Kaulakys}(1991)}]{1991JPhB...24L.127K}
{Kaulakys}, B. 1991, Journal of Physics B Atomic Molecular Physics, 24, L127

\bibitem[{{Korn} {et~al.}(2003){Korn}, {Shi}, \&
  {Gehren}}]{2003A&A...407..691K}
{Korn}, A.~J., {Shi}, J., \& {Gehren}, T. 2003, \aap, 407, 691

\bibitem[{{Kroll} \& {Kock}(1987)}]{KK}
{Kroll}, S. \& {Kock}, M. 1987, Astron. and Astrophys. Suppl. Ser., 67, 225,
  (KK)

\bibitem[{{Kurucz}(1992)}]{1992RMxAA..23...45K}
{Kurucz}, R.~L. 1992, \rmxaa, 23

\bibitem[{{Kurucz}(2009)}]{Kurucz2009}
{Kurucz}, R.~L. 2009, Robert L. Kurucz on-line database of observed and
  predicted atomic transitions

\bibitem[{{Lind} {et~al.}(2017){Lind}, {Amarsi}, {Asplund}, {Barklem},
  {Bautista}, {Bergemann}, {Collet}, {Kiselman}, {Leenaarts}, \&
  {Pereira}}]{2017MNRAS.468.4311L}
{Lind}, K., {Amarsi}, A.~M., {Asplund}, M., {et~al.} 2017, \mnras, 468, 4311

\bibitem[{{Lind} {et~al.}(2012){Lind}, {Bergemann}, \&
  {Asplund}}]{2012MNRAS.427...50L}
{Lind}, K., {Bergemann}, M., \& {Asplund}, M. 2012, \mnras, 427, 50

\bibitem[{{Mashonkina} {et~al.}(2011){Mashonkina}, {Gehren}, {Shi}, {Korn}, \&
  {Grupp}}]{mash_fe}
{Mashonkina}, L., {Gehren}, T., {Shi}, J.-R., {Korn}, A.~J., \& {Grupp}, F.
  2011, \aap, 528, A87

\bibitem[{{Mashonkina} {et~al.}(2017{\natexlab{a}}){Mashonkina}, {Jablonka},
  {Pakhomov}, {Sitnova}, \& {North}}]{dsph_parameters}
{Mashonkina}, L., {Jablonka}, P., {Pakhomov}, Y., {Sitnova}, T., \& {North}, P.
  2017{\natexlab{a}}, \aap, 604, A129

\bibitem[{{Mashonkina} {et~al.}(2017{\natexlab{b}}){Mashonkina}, {Jablonka},
  {Sitnova}, {Pakhomov}, \& {North}}]{2017A&A...608A..89M}
{Mashonkina}, L., {Jablonka}, P., {Sitnova}, T., {Pakhomov}, Y., \& {North}, P.
  2017{\natexlab{b}}, \aap, 608, A89

\bibitem[{{Mel{\'e}ndez} \& {Barbuy}(2009)}]{MB09}
{Mel{\'e}ndez}, J. \& {Barbuy}, B. 2009, \aap, 497, 611

\bibitem[{{Moity}(1983)}]{1983A&AS...52...37M}
{Moity}, J. 1983, \aaps, 52, 37

\bibitem[{{Nave} {et~al.}(1994){Nave}, {Johansson}, {Learner}, {Thorne}, \&
  {Brault}}]{1994ApJS...94..221N}
{Nave}, G., {Johansson}, S., {Learner}, R.~C.~M., {Thorne}, A.~P., \& {Brault},
  J.~W. 1994, \apjs, 94, 221

\bibitem[{{O'Brian} {et~al.}(1991){O'Brian}, {Wickliffe}, {Lawler}, {Whaling},
  \& {Brault}}]{1991JOSAB...8.1185O}
{O'Brian}, T.~R., {Wickliffe}, M.~E., {Lawler}, J.~E., {Whaling}, W., \&
  {Brault}, J.~W. 1991, Journal of the Optical Society of America B Optical
  Physics, 8, 1185

\bibitem[{{Pakhomov} {et~al.}(2019){Pakhomov}, {Mashonkina}, {Sitnova}, \&
  {Jablonka}}]{Pakhomov_boo1}
{Pakhomov}, Y., {Mashonkina}, L., {Sitnova}, T., \& {Jablonka}, P. 2019,
  Astronomy Letters, 45, 303

\bibitem[{{Pauls} {et~al.}(1990){Pauls}, {Grevesse}, \& {Huber}}]{PGHcor}
{Pauls}, U., {Grevesse}, N., \& {Huber}, M.~C.~E. 1990, Astron. and Astrophys.,
  231, 536

\bibitem[{{Raassen} \& {Uylings}(1998)}]{RU}
{Raassen}, A.~J.~J. \& {Uylings}, P.~H.~M. 1998, \aap, 340, 300

\bibitem[{{Roederer} {et~al.}(2018){Roederer}, {Sneden}, {Lawler}, {Sobeck},
  {Cowan}, \& {Boesgaard}}]{2018ApJ...860..125R}
{Roederer}, I.~U., {Sneden}, C., {Lawler}, J.~E., {et~al.} 2018, \apj, 860, 125

\bibitem[{{Ruchti} {et~al.}(2013){Ruchti}, {Bergemann}, {Serenelli},
  {Casagrande}, \& {Lind}}]{Ruchti2013}
{Ruchti}, G.~R., {Bergemann}, M., {Serenelli}, A., {Casagrande}, L., \& {Lind},
  K. 2013, \mnras, 429, 126

\bibitem[{{Ruffoni} {et~al.}(2014){Ruffoni}, {Den Hartog}, {Lawler}, {Brewer},
  {Lind}, {Nave}, \& {Pickering}}]{2014MNRAS.441.3127R}
{Ruffoni}, M.~P., {Den Hartog}, E.~A., {Lawler}, J.~E., {et~al.} 2014, \mnras,
  441, 3127

\bibitem[{{Ryabchikova} {et~al.}(2015){Ryabchikova}, {Piskunov}, {Kurucz},
  {Stempels}, {Heiter}, {Pakhomov}, \& {Barklem}}]{2015PhyS...90e4005R}
{Ryabchikova}, T., {Piskunov}, N., {Kurucz}, R.~L., {et~al.} 2015, \physscr,
  90, 054005

\bibitem[{{Seaton}(1962)}]{1962amp..conf..375S}
{Seaton}, M.~J. 1962, in Atomic and Molecular Processes, ed. D.~R. {Bates}, 375

\bibitem[{{Shchukina} \& {Trujillo Bueno}(2001)}]{2001ApJ...550..970S}
{Shchukina}, N. \& {Trujillo Bueno}, J. 2001, \apj, 550, 970

\bibitem[{{Sitnova} {et~al.}(2015){Sitnova}, {Zhao}, {Mashonkina}, {Chen},
  {Liu}, {Pakhomov}, {Tan}, {Bolte}, {Alexeeva}, {Grupp}, {Shi}, \&
  {Zhang}}]{2015ApJ...808..148S}
{Sitnova}, T., {Zhao}, G., {Mashonkina}, L., {et~al.} 2015, \apj, 808, 148

\bibitem[{{Sitnova} {et~al.}(2019){Sitnova}, {Mashonkina}, {Ezzeddine}, \&
  {Frebel}}]{2019MNRAS.485.3527S}
{Sitnova}, T.~M., {Mashonkina}, L.~I., {Ezzeddine}, R., \& {Frebel}, A. 2019,
  \mnras, 485, 3527

\bibitem[{{Sneden} {et~al.}(2016){Sneden}, {Cowan}, {Kobayashi}, {Pignatari},
  {Lawler}, {Den Hartog}, \& {Wood}}]{2016ApJ...817...53S}
{Sneden}, C., {Cowan}, J.~J., {Kobayashi}, C., {et~al.} 2016, \apj, 817, 53

\bibitem[{{Steenbock} \& {Holweger}(1984)}]{Steenbock1984}
{Steenbock}, W. \& {Holweger}, H. 1984, \aap, 130, 319

\bibitem[{{Tanaka}(1971)}]{1971PASJ...23..217T}
{Tanaka}, K. 1971, \pasj, 23, 217

\bibitem[{{Tsymbal} {et~al.}(2019){Tsymbal}, {Ryabchikova}, \&
  {Sitnova}}]{2019ASPC}
{Tsymbal}, V., {Ryabchikova}, T., \& {Sitnova}, T. 2019, in Astronomical
  Society of the Pacific Conference Series, Vol. 518, Astronomical Society of
  the Pacific Conference Series, ed. D.~O. {Kudryavtsev}, I.~I. {Romanyuk}, \&
  I.~A. {Yakunin}, 247--252

\bibitem[{{Yakovleva} {et~al.}(2018{\natexlab{a}}){Yakovleva}, {Barklem}, \&
  {Belyaev}}]{2018MNRAS.473.3810Y}
{Yakovleva}, S.~A., {Barklem}, P.~S., \& {Belyaev}, A.~K. 2018{\natexlab{a}},
  \mnras, 473, 3810

\bibitem[{{Yakovleva} {et~al.}(2018{\natexlab{b}}){Yakovleva}, {Belyaev}, \&
  {Kraemer}}]{2018CP....515..369Y}
{Yakovleva}, S.~A., {Belyaev}, A.~K., \& {Kraemer}, W.~P. 2018{\natexlab{b}},
  Chemical Physics, 515, 369

\bibitem[{{Yakovleva} {et~al.}(2019){Yakovleva}, {Belyaev}, \&
  {Kraemer}}]{2019MNRAS.483.5105Y}
{Yakovleva}, S.~A., {Belyaev}, A.~K., \& {Kraemer}, W.~P. 2019, \mnras, 483,
  5105

\bibitem[{{Zhang} \& {Pradhan}(1995)}]{1995A&A...293..953Z}
{Zhang}, H.~L. \& {Pradhan}, A.~K. 1995, \aap, 293, 953

\bibitem[{{Zhao} {et~al.}(2016){Zhao}, {Mashonkina}, {Yan}, {Alexeeva},
  {Kobayashi}, {Pakhomov}, {Shi}, {Sitnova}, {Tan}, {Zhang}, {Zhang}, {Zhou},
  {Bolte}, {Chen}, {Li}, {Liu}, \& {Zhai}}]{lick_paperII}
{Zhao}, G., {Mashonkina}, L., {Yan}, H.~L., {et~al.} 2016, \apj, 833, 225

\end{thebibliography}
\bibliographystyle{aa}

\end{document}